\pdfoutput=1
\documentclass[acmsmall,usenames,dvipsnames,table,screen]{acmart}

\usepackage{epsf}
\usepackage{epsfig}
\usepackage{graphicx}
\usepackage{ifthen}
\usepackage{comment}
\usepackage{url}
\usepackage{hyperref}
\usepackage{latexsym}

\usepackage{amsfonts}
\usepackage{subfigure}
\usepackage{amsmath}
\usepackage{url}
\usepackage{multirow}
\usepackage{floatflt}
\usepackage{listings}
\usepackage{icomma}
\PassOptionsToPackage{usenames,dvipsnames}{color}
\PassOptionsToPackage{table}{xcolor}

\usepackage{xspace}
\usepackage{enumitem}
\usepackage{tabularx}
\usepackage{makecell}
\usepackage{tikz}
\usepackage[normalem]{ulem}

\usepackage{tabularx}
\usepackage{booktabs}

\usepackage{csquotes}

\newcommand{\tool}{$\mu$SE\xspace}
\newcommand{\tools}{$\mu$SE's\xspace}
\newcommand{\mdroid}{{\small MDroid$+$}\xspace}

\newboolean{showcomments}

\setboolean{showcomments}{true}

\ifthenelse{\boolean{showcomments}}
  {\newcommand{\nb}[2]{
    \fbox{\bfseries\sffamily\scriptsize#1}
    {\sf\small$\blacktriangleright$\textit{#2}$\blacktriangleleft$}
   }
   
  }
  {\newcommand{\nb}[2]{}
   
  }

\newcommand\myparagraph[1]{\noindent {\bf {#1}:}}

\newcommand{\remove}[1]{{\leavevmode{\xspace#1}}}
\newcommand{\add}[1]{{\leavevmode\color{blue}{\xspace#1}}}

\newcommand{\ie}{\textit{i.e.}\xspace}
\newcommand{\eg}{\textit{e.g.}\xspace}

\newcommand{\etal}{\textit{et al.}\xspace}
\newcommand{\etals}{\textit{et al.'s}\xspace}

\newcommand{\recycler}{\textsf{RecyclerView}\xspace}
\newcommand{\soundiness}{{sound{\em i}ness}\xspace}
\newcommand{\recycleradapter}{\textsf{RecyclerView.Adapter}\xspace}
\newcommand{\recyclerviewholder}{\textsf{RecyclerView.ViewHolder}\xspace}
\newcommand{\reachability}{{\textsf{Reachability}\xspace}}
\newcommand{\complexreachability}{{\textsf{Complex Reachability}\xspace}}
\newcommand{\taint}{{\textsf{Taint}\xspace}}
\newcommand{\scope}{{\textsf{Scope}\xspace}}

\newcommand{\countinsertedtotalleaks}{{\(54,936\)}\xspace}
\newcommand{\countexecutabletotalleak}{{\(8,250\)\xspace}}
\newcommand{\countnonexecutabletotalleak}{{\(46,686\)\xspace}}

\newcommand{\countinsertoldleak}{{\(7,584\)}\xspace}
\newcommand{\countinsertoldexecutableleak}{{\(2,026\)}\xspace}
\newcommand{\countinsertnewleak}{{\(30,117\)}\xspace}
\newcommand{\countexecutablenewleak}{{\(4,385\)}\xspace}

\newcommand{\countnonexecutabletotalleakpercentage}{{\(85\)\xspace}}

\newcommand{\counttotalhours}{{\(19\)}\xspace}

\newcommand{\countmutantappsfromlatereight}{{\(31\)}\xspace}

\newcommand{\flowdroidnotdetected}{{\(987\)}\xspace}

\newcommand{\horndroidtestedagainst}{{\(46\)}\xspace}
\newcommand{\horndroiddetected}{{\(14\)}\xspace}
\newcommand{\horndroidnotdetected}{{\(32\)}\xspace}

\newcommand{\argusnotdetected}{{\(7,708\)}\xspace}

\newcommand{\usenix}{{USENIX'18}\xspace}

\definecolor{dkgreen}{rgb}{0,0.6,0}
\definecolor{gray}{rgb}{0.5,0.5,0.5}
\definecolor{mauve}{rgb}{0.58,0,0.82}
\definecolor{shadecolor}{rgb}{0.95,0.95,0.95}

\definecolor{pblue}{rgb}{0.13,0.13,1}
\definecolor{pgreen}{rgb}{0,0.5,0}
\definecolor{pred}{rgb}{0.9,0,0}
\definecolor{pgrey}{rgb}{0.46,0.45,0.48}

\lstdefinestyle{toplisting}{
  float=tp,
  floatplacement=tbp,
}

\AtBeginDocument{%
  \providecommand\BibTeX{{%
    \normalfont B\kern-0.5em{\scshape i\kern-0.25em b}\kern-0.8em\TeX}}}

\setcopyright{acmcopyright}
\acmJournal{TOPS}
\acmYear{2020} \acmVolume{1} \acmNumber{1} \acmArticle{1} \acmMonth{1} \acmPrice{15.00}
\acmDOI{10.1145/3439802}

\begin{document}

\title[Evaluation of Android Security using Mutation]{Systematic
Mutation-based Evaluation \\ of the Soundness of Security-focused \\Android Static Analysis Techniques}

\author{Amit Seal Ami}
\email{aami@email.wm.edu}
\orcid{0002-9455-2230}
\affiliation{%
  \institution{William \& Mary, Department of Computer Science}
  \streetaddress{P.O.Box 8795}
  \city{Williamsburg}
  \state{VA}
  \postcode{23185}
}

\author{Kaushal Kafle}
\email{kkafle@email.wm.edu}
\affiliation{%
  \institution{William \& Mary, Department of Computer Science}
  \streetaddress{P.O.Box 8795}
  \city{Williamsburg}
  \state{VA}
  \postcode{23185}
}

%\author{Richard Bonett}
%\email{rfbonett@email.wm.edu}
\author{Kevin Moran}
\email{kpmoran@gmu.edu}
\affiliation{%
  \institution{George Mason University, Department of Computer Science}
  \streetaddress{4400 University Dr}
  \city{Fairfax}
  \state{VA}
  \postcode{22030}
}

\author{Adwait Nadkarni}
\email{nadkarni@cs.wm.edu}
\affiliation{%
  \institution{William \& Mary, Department of Computer Science}
  \streetaddress{P.O.Box 8795}
  \city{Williamsburg}
  \state{VA}
  \postcode{23185}
}

\author{Denys Poshyvanyk}
\email{denys@cs.wm.edu}
\affiliation{%
  \institution{William \& Mary, Department of Computer Science}
  \streetaddress{P.O.Box 8795}
  \city{Williamsburg}
  \state{VA}
  \postcode{23185}
}

\authorsaddresses{%
  Author's addresses: Amit Seal Ami, aami@email.wm.edu; Kaushal Kafle, kkafle@cs.wm.edu; Adwait Nadkarni, nadkarni@cs.wm.edu; Denys Poshyvanyk, denys@cs.wm.edu: Computer Science Department, Williamsburg, VA, USA. Kevin Moran, kpmoran@gw.edu: Computer Science  Department, George Mason University, Fairfax, VA, USA}

\renewcommand{\shortauthors}{Amit Seal Ami, Kaushal Kafle, Kevin Moran, et al.}

\begin{abstract}
% Area
Mobile application security has been a major area of focus for security research over the course of the last decade. Numerous application analysis tools have been proposed in response to malicious, curious, or vulnerable apps.
% Problem
However, existing tools, and specifically, static analysis tools, trade soundness of the analysis for precision and performance and are hence sound{\em y}. Unfortunately, the specific unsound choices or flaws in the design of these tools is often not known or well-documented, leading to misplaced confidence among researchers, developers, and users.
% Solution
This paper describes the {\em Mutation-based Soundness Evaluation} (\tool)
framework, which systematically evaluates Android static analysis tools to discover, document, and fix flaws, by leveraging the well-founded practice of mutation analysis.
% Methodology
We implemented \tool{} and applied it to a set of prominent Android static analysis tools that detect private data leaks in apps. In a study conducted previously, we used \tool to discover $13$ previously undocumented flaws in FlowDroid, one of the most prominent data leak detectors for Android apps.
Moreover, we discovered that flaws also propagated to other tools that build upon the design or implementation of FlowDroid or its components.
This paper substantially extends our \tool framework and offers an new in-depth analysis of two more major tools in our 2020 study, we find $12$ new, undocumented flaws and demonstrate that all $25$ flaws are found in more than one tool, regardless of any inheritance-relation among the tools.
%Takeaway
Our results motivate the need for systematic discovery and documentation of unsound choices in sound{\em y} tools and demonstrate the opportunities in leveraging mutation testing in achieving this goal.
\end{abstract}




\maketitle

\section{Introduction}
\label{sec:intro}

Mobile devices such as smartphones and tablets have become the fabric of our consumer computing ecosystems; by the year \(2020\), more than \(80\%\) of the world's adult population is projected to own a smartphone~\cite{economist}. This popularity of mobile devices is driven by the millions of diverse, feature-rich, third-party applications or ``apps'' they support. However, in fulfilling their functionality, apps often require access to security and privacy-sensitive resources on the device (\eg{}  GPS location, security settings). Applications can neither be trusted to be well-written or benign, and to prevent misuse of such access through malicious or vulnerable apps~\cite{lnw+14,gzz+12,zj12,rce13,fhm+12,ssg+14,ebfk13}, it is imperative to understand the challenges in securing mobile apps.

Security analysis of third-party apps has been one of the dominant areas of smartphone security research in the last decade, resulting in tools and frameworks with diverse security goals. For instance, prior work has designed tools to identify malicious behavior in apps~\cite{eom09b,zwzj12,ash+14}, discover private data leaks~\cite{egc+10,arf+14,gcec12,akg+15}, detect vulnerable application interfaces~\cite{fwm+11,cfgw11,llw+12,lbs+17}, identify flaws in the use of cryptographic primitives~\cite{fhm+12,ebfk13,ssg+14}, and define sandbox policies for third-party apps~\cite{hdge14,jsz16}. To protect users from malicious or vulnerable apps, it is imperative to assess the challenges and pitfalls of existing tools and techniques. However, {\em it is unclear whether existing security tools are sufficiently robust to expose particularly well-hidden unwanted behaviors.}

Our work is motivated by the pressing need to discover the limitations of application analysis techniques for Android. Existing application analysis techniques, specifically those that employ static analysis, must in practice trade soundness for precision, as there is an inherent conflict between the two properties. A sound analysis requires the technique to over-approximate (\ie{} consider instances of unwanted behavior that may not execute in reality), which in turn deteriorates precision. This trade-off has practical implications on the security provided by static analysis tools.  That is, {\em in theory}, static analysis is expected to be sound, yet, in practice, these tools must purposefully make unsound choices to achieve a feasible analysis that has sufficient precision and performance to scale. For instance, techniques that analyze Java generally do not over-approximate analysis of certain programming language features, such as reflection, for practical reasons (\eg{} Soot~\cite{vcg+99}, FlowDroid~\cite{arf+14}). Although this particular case is well-known and documented, many such unsound design choices are neither well-documented, nor known to researchers outside a small community of experts.

Security experts have described such tools as sound{\em y}, \ie{} having a core set of sound design choices, in addition to certain practical assumptions that sacrifice soundness for precision~\cite{lss+15}. Although soundness is an elusive ideal, sound{\em y} tools certainly seem to be a practical choice: {\em but only if the unsound choices are known, necessary, and properly documented}. However, the present state of sound{\em y} static analysis techniques is dire, as unsound choices {\sf (1)}~may not be documented, and unknown to non-experts, {\sf (2)}~may not even be known to tool designers (\ie{} implicit assumptions), and {\sf (3)}~may propagate into future research.  The \soundiness{} manifesto describes the misplaced confidence generated by the insufficient study and documentation of sound{\em y} tools, in the specific context of language features~\cite{lss+15}. Motivated by the manifesto, we leverage \soundiness{} at the general, conceptual level of design choices, and attempt to resolve the status quo of sound{\em y} tools by making them more secure as well as transparent.

We describe the {\em Mutation-based Soundness Evaluation} (\tool{}, read as ``muse'') framework that enables systematic security evaluation of Android static analysis tools to discover unsound design assumptions, leading to their documentation, as well as improvements in the tools themselves.
% {\color{blue}
In particular, this manuscript describes our extension of our original \tool{} paper published in \usenix{} in August~\cite{richie18}.
% }
\tool{} leverages the practice of mutation analysis from the software engineering (SE) domain~\cite{Offutt2001, Hamlet:TSE,DeMillo:Computer, Ma:ISSRE03,Derezinska2014, Praphamontripong:Mutation15, Appelt:ISSTA14, ZhouF09, Oliveira:ICSTW15, Nardo:ICST15}, and specifically, more recent advancements in mutating Android apps~\cite{lbt+17}.
In doing so, \tool{} adapts a well-founded practice from SE to security, by making useful changes to contextualize it to evaluate security tools.

\tool{} creates {\em security operators}, which reflect the security goals of the tools being analyzed (\eg{} data leak or SSL vulnerability detection). These security operators are seeded, \ie{} inserted into one or more Android apps, as guided by a {\em mutation scheme}. This seeding results in the creation of multiple mutants (\ie{} code that represents the target unwanted behavior) within the app. Finally, the mutated application is analyzed using the security tool being evaluated, and the undetected mutants are then subjected to a deeper analysis. We propose a semi-automated methodology to analyze the uncaught mutants, resolve them to flaws in the tool, and confirm the flaws experimentally.

We demonstrate the effectiveness of our approach by evaluating static analysis research tools that detect data leaks in Android apps.
Based on our analysis of the discovered flaws, we provide immediate patches that address one flaw, and identify classes of design-level flaws that may be hard to address without significant research effort.
Further, we perform a flaw propagation study that checks for the presence of these flaws among {\em seven} data leak detectors.

The general contributions of this paper can be summarized as follows:

\begin{itemize}\renewcommand{\itemsep}{-0.1em}
  \item {\em We introduce the novel paradigm of Mutation-based Soundness  Evaluation (\ie\ \tool)}, which provides a systematic methodology for discovering flaws in   static analysis tools for Android, leveraging the well-understood practice of   mutation analysis.  We adapt mutation analysis for security evaluation and design the abstractions of {\em security operators} and {\em mutation   schemes}.
  \item {\em We design and implement the \tool{} framework} for evaluating    Android static analysis tools. \tool{} adapts to the security goals of a tool being evaluated and allows the detection of unknown or undocumented flaws.
  \item {\em We demonstrate the effectiveness of \tool{} by evaluating several widely used Android security tools} that detect private data leaks in Android   apps.  \tool{} detects
  % \(25\)
  undocumented flaws, and demonstrates their propagation.   Our analysis leads to the documentation of unsound assumptions at the design-level, and immediate security patches for an easily fixable but evasive flaw.

\end{itemize}
% \add{
We published an earlier version of this work in~\usenix~\cite{richie18} where we analyzed FlowDroid~\cite{arf+14} using \tool{} and discovered $13$ previously undocumented flaws. The current version of this study substantially extends upon the previous work, in the following manner:

\begin{itemize}
  \item {\bf Design and Implementation of \tool:} We designed a new scope-based mutation placement approach, in addition to the mutation schemes originally explored in the \usenix paper~\cite{richie18}. Further, we refined the implementation of the reachability-based scheme by integrating class declaration-level placement. Moreover, this extension includes a discussion on leveraging the abstractions invented in \tool for evaluating tools with security goals other than data leak detection, such as the detection of cryptographic-API misuse vulnerabilities. Such a discussion on the general applicability of \tools abstractions was not present in our \usenix paper. Finally, we enhance \tools execution engine as a part of this extension, thereby improving its accuracy in terms of associating execution traces with mutants.
  \item {\bf Multiplicative improvement in mutants generated:} In the USENIX'18 paper~\cite{richie18}, 7 Android apps were used as base applications for mutation, to produce $7,584$ compilable mutants. In this extension, we added 8 new apps, resulting in a total of 15 real-world, open-source Android apps used for mutation.
  We were able to seed $30,117$ compilable mutants into these $8$ new apps, of which \countexecutablenewleak were executable, and were used for evaluating additional data leak detectors (see next).
  More importantly, to gauge the impact of our design and implementation improvements on \tools ability to generate compilable mutants, we also mutated the $7$ original apps, which resulted in $24,819$ mutants, \ie, $17,235$ more mutants (a {\em $2.27x$ increase in compilable mutants}) over the USENIX'18 study.
  \item{\bf Evaluation of the Effectiveness of Mutation Schemes:} In this extension, we perform a data-driven evaluation of our implemented mutation schemes, across two primary directions: {\sf (1)} their ability to create executable mutants, and {\sf (2)} their ability to create mutants that may lead to the discovery of flaws.
 Our analysis, and the resultant insights regarding schemes is novel, as no such evaluation was performed in our \usenix paper~\cite{richie18}.

  \item {\bf Significant additional extrinsic evaluation with an in-depth analysis of Argus and HornDroid:} We studied FlowDroid in-depth in our USENIX'18 paper, which formed {\em the core of the extrinsic evaluation}. In this study, we effectively {\em tripled our extrinsic evaluation} by performing an in-depth soundness evaluation of two additional state-of-the-art data leak detecting static analysis approaches for Android, namely Argus (previously known as AmanDroid)~\cite{wror14} and HornDroid~\cite{cgm16}. We used the newly generated set of \countexecutablenewleak{} executable mutants generated from the 8 new base apps to evaluate Argus and HornDroid.
  % New flaws found, new flaw class
  \item {\bf New Findings:} Our extended analysis led to significant new findings over our \usenix study~\cite{richie18}, which can be summarized into the following four points. {\sf (1)} {\bf New flaws.} We found $12$ novel flaws in Argus and HornDroid, in addition to the $13$ flaws discovered in FlowDroid in our USENIX'18 study~\cite{richie18}, which brings the total number of discovered flaws to $25$.
  {\sf (2)} {\bf New flaw class.} We discovered a {\em new flaw class}, aside from the four classes discussed in the USENIX'18 version, due to a distinctive and novel pattern exhibited by certain flaws, \ie, a fundamentally flawed analysis of critical Android lifecycle methods that are present in all applications (e.g., onCreate).
 {\sf (3)} {\bf New insight on propagation.} Furthermore, we discovered a {\em new insight in terms of how flaws propagate across tools}, relative to the USENIX'18 paper.
  That is, in the USENIX paper we found that flaws generally propagate when a direct inheritance relationship is present among two tools (\ie, directly relying on the code base), but also observed that the flaws did not propagate to tools that did not have a direct relationship, but were simply built for similar design goals.
  However, in this work, on studying the propagation of flaws with four additional tools (\ie, across total $7$ tools), we discovered that every single flaw was present in at least one other tool, which means that flaws propagate across tools purely because of the shared design goal, even without any direct inheritance relationship.
  Such propagation based purely on the design goal was speculated in the USENIX'18 paper, but was not evident, as it is in this work.
 {\sf (4)} {\bf New insight on re-emergence of flaws.} Finally, we found that flaws that are fixed after reporting, {\em can re-emerge in future updates} of a tool.
  Our findings demonstrate that soundness issues can be introduced at any point in tools' lifecycles, which further necessitates continuous evaluation with \tool.
\end{itemize}

% }
% \add{
We have released the \tool framework, the security operators and mutation schemes constructed for evaluating data leak detectors, as well as all of the experimental data, to facilitate the reproducibility of our results, as well as to enable  security researchers, tool designers, and analysts uncover undocumented flaws and unsound choices in sound{\em y} security tools~\cite{appendix}.

\section{Motivation and Background}
\label{sec:motivation}

This work is motivated by the pressing need to help researchers and  practitioners identify instances of unsound assumptions or design decisions in their static analysis tools, thereby {\em extending the sound core} of their sound{\em y} techniques.  That is, security tools may already have a core set of sound design decisions (\ie{} the sound core) and may claim soundness  based on those decisions. The \soundiness{} manifesto~\cite{lss+15} defines the {\em sound core} in terms of specific language features, whereas we use the term in a more abstract manner to refer to the design goals of the tool. Systematically identifying unsound decisions may allow researchers to resolve flaws and help extend the sound core of their tools.

Moreover, research papers and tool documentations indeed do not articulate many  of the unsound assumptions and design choices that lie outside their sound core, aside from some well-known cases (\eg{} choosing not to handle reflection, race conditions), as confirmed by our results (Section~\ref{sec:eval}). In addition, there is a chance that developers of these techniques may be unaware of some implicit assumptions/flaws due to a host of reasons: \eg{}  because the assumption was inherited from prior research or a certain aspect of Android was not modeled correctly. Therefore, our objective is to discover instances of such hidden assumptions and design flaws that affect the security claims made by tools, document them explicitly, and possibly, help developers mend existing artifacts.

\subsection{Motivating Example}
\label{sec:example}
Consider the following motivating example of a prominent static analysis tool,  FlowDroid~\cite{arf+14}:\\ FlowDroid is a highly popular static analysis framework for detecting private data leaks in Android apps by performing a data flow analysis.  Some of the limitations of FlowDroid are well-known and stated in the paper~\cite{arf+14}; \eg{} FlowDroid does not support reflection, like most static analyses for Java. However, through a manual but systematic, preliminary, analysis of FlowDroid, we discovered a security limitation that is not well-known or accounted for in the paper, and hence affects guarantees provided by the tool's analysis.
We discovered that FlowDroid (\ie{} v1.5, which was latest at the time of our preliminary analysis in October 2017) does not support ``Android fragments''~\cite{android-fragments}, which are app modules that are widely used in most Android apps (\ie{} in more than 90\% of the top 240 Android apps per category on Google Play, as demonstrated in our original \usenix paper~\cite{richie18}).
This flaw renders any security analysis of general Android apps using FlowDroid unsound, due to the high likelihood of fragment use, even when the app developers may be cooperative and non-malicious. Further, FlowDroid v2.0, which was released on 10/10/2017~\cite{fdroid_release}, claims to address fragments, {\em but failed to detect our exploit}.

On investigating further, we found that FlowDroid v1.5 was extended or used by at least 13 research tools~\cite{lbb+15, kfb+14, Yang:2015, akg+15, Octeau:2015, Sasnauskas:2014, Liu:2015, Slavin:2016, Aafer:2015, Rasthofer:2014, Li:2014, Lillack:2017, Nan:2015},
none of which acknowledge or address this limitation in modeling fragments. This  leads us to conclude that this significant flaw not only persists in FlowDroid, but may have additionally propagated to the tools that inherit it directly (\ie, by inheriting its code), or indirectly (\ie, by adhering to similar design principles/goals).
Our flaw propagation study in Section~\ref{sec:propagation_study_pre} confirms this conjecture for inheritors of FlowDroid, as well as the two other data leak detectors that we evaluate in-depth, \ie, Argus~\cite{wror14} and Horndroid~\cite{cgm16}.

We reported the flaws to the authors of FlowDroid, and created two  patches to fix it.
Our patches were confirmed to work on FlowDroid v\(2.0\) built from source, and  were accepted into FlowDroid's repository~\cite{fdroid_git} in December 2017.
Thus, we were able to discover and fix an undocumented flaw that significantly affected FlowDroid's soundness claims, thereby expanding its sound core.  However, we have confirmed that FlowDroid v\(2.5\)~\cite{fdroid_release} still fails to detect leaks in fragments, and are working with developers to resolve this issue.

Through this example, we demonstrate that unsound assumptions in  security-focused static analysis tools for Android are not only detrimental to the validity of their own analysis, but may inadvertently propagate to future research.  Thus, identifying these unsound assumptions is not only beneficial for making the user of the analysis aware of its true limits, but in addition for the research community in general.  As of today, aside from a handful of manually curated testing tool-kits (\eg, DroidBench~\cite{arf+14}) with hard-coded (but useful) checks, to the best of our knowledge, there has been no prior effort at methodologically discovering problems related to \soundiness in Android static analysis tools and frameworks.  {\em This paper is motivated by the need to systematically identify and resolve the unsound assumptions in security-related static analysis tools.}

\subsection{Background on Mutation Analysis}
\label{subsec:background}

Mutation analysis has a strong foundation in the field of SE, and is typically  used  as a test adequacy criterion, measuring the effectiveness of a set of test cases~\cite{Offutt2001}. Faulty programs are created by applying transformation rules, called \textit{mutation operators} to a given program. The larger the number of faulty programs or \textit{mutants} detected by a test suite, the higher the effectiveness of that particular suite.
Since its inception~\cite{Hamlet:TSE,DeMillo:Computer}, mutation testing has  seen striking advancements related to the design and development of advanced operators. Research related to development of mutation operators has traditionally attempted to adapt operators for a particular target domain, such as the web~\cite{Praphamontripong:Mutation15}, data-heavy applications~\cite{Appelt:ISSTA14,ZhouF09,Nardo:ICST15}, or GUI-centric applications~\cite{Oliveira:ICSTW15}.
Recently, mutation analysis has been applied to measure the effectiveness of  test suites for both functional and non-functional requirements of Android apps~\cite{Deng:ICSTW15,Jabbarvand:FSE17, lbt+17}.

This paper builds upon SE concepts of mutation analysis and adapts them to a  security context. Our methodology does not simply use the traditional mutation analysis, but rather {\em redefines} it to effectively improve security-focused static analysis tools, as described in Sections~\ref{sec:design} and ~\ref{sec:implementation}.

\section{\tool}
\label{sec:overview}

\begin{figure}[t]
    \centering
    \includegraphics[width=.55\linewidth]{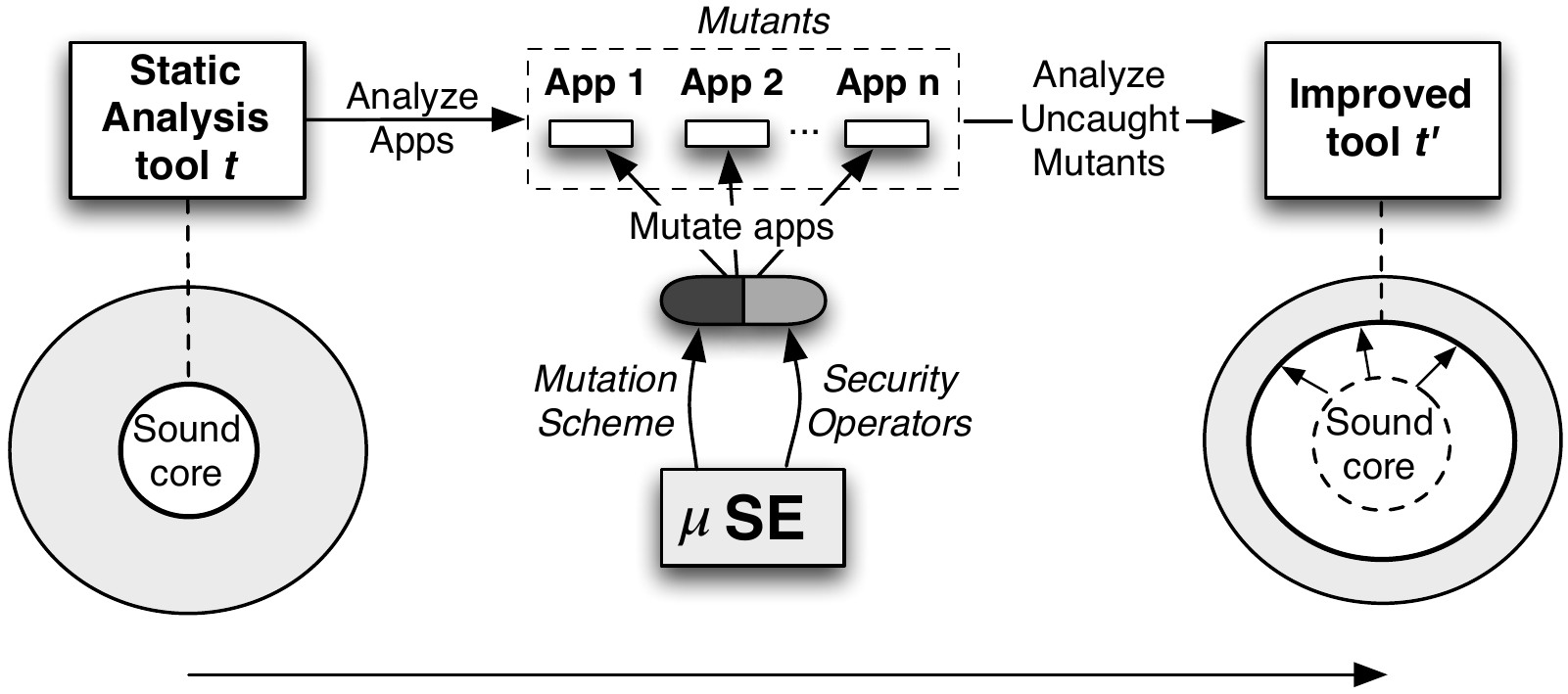}
     \vspace{-1em}
    \caption{{\small \tool{} tests a static analysis tool on a set of mutated
    Android apps and analyzes uncaught mutants to discover and/or fix flaws.}}
    \Description[Architecture of \tool]{Shows interaction between \tool, apps
    and static analysis tools}
    \vspace{-1em}
    \label{fig:expanding_core}
\end{figure}
We describe \tool, a semi-automated framework for systematically evaluating  Android static analysis tools that adapts the process of mutation analysis commonly used to evaluate software test suites~\cite{Offutt2001}. That is, we aim to help discover concrete instances of flawed security design decisions made by static analysis tools, by exposing them to methodologically mutated applications. We envision two primary benefits from \tool: {\em short-term} benefits related to straightforwardly fixable flaws that may be patched immediately, and  {\em long-term} benefits related to the continuous documentation of assumptions and flaws, even those that may be hard to resolve. This section provides an overview of \tool (Figure~\ref{fig:expanding_core}) and its design goals.

 As shown in Figure~\ref{fig:expanding_core}, we take an Android static analysis  tool to be evaluated (\eg, FlowDroid~\cite{arf+14} or MalloDroid~\cite{fhm+12}) as input. \tool executes the tool on {\em mutants}, \ie, apps to which {\em security operators} (\ie, security-related mutation operators) are applied, as per a \textit{mutation scheme}, which governs the placement of code transformations described by operators in the app (\ie, thus generating mutants). The security operators represent anomalies that the static analysis tools are expected to detect, and hence, are closely tied to the security goal of the tool. The uncaught mutants indicate flaws in the tool, and analyzing them leads to the broader discovery and awareness of the unsound assumptions of the tools, eventually facilitating security-improvements.

\myparagraph{Design Goals} Measuring the security provided by a system is a difficult problem; however, we may be able to better predict failures if the assumptions made by the system are known in advance. Similarly, although soundness may be a distant ideal for security tools, we assert that it should be feasible to articulate the boundaries of a tool's sound core. Knowing these boundaries would be immensely useful for analysts who use security tools, for developers looking for ways to improve tools, as well as for end users who benefit from the security analyses provided by such tools. To this end, we design \tool to provide an effective foundation for evaluating Android security tools. Our design of \tool is guided by the following goals:
\begin{enumerate}[label=\textbf{G\arabic*},ref=\textbf{G\arabic*}]

  \item\label{goal:operator} {\em \textbf{Contextualized security operators.}} Android security tools have diverse purposes and may claim various security guarantees. Security operators must be instantiated in a way that is sensitive to the context or purpose (\eg, data leak identification) of the tool being evaluated.
  \item\label{goal:android} {\em  \textbf{Android-focused mutation schemes.}} Android's security challenges are notably unique, and hence require a diverse array of novel security analyses.  Thus, the strategies for defining mutation schemes, \ie, the {\em placement} of the target, unwanted behavior in the app, must consider Android's abstractions and application model for effectiveness.
  \item\label{goal:feasibility} {\em \textbf{Minimize manual-effort during analysis.}} Although \tool{} is certainly more feasible than manual analysis, we intend to  significantly reduce the manual effort spent on evaluating undetected mutants. Thus, our goal is to dynamically filter inconsequential mutants, and to develop a systematic methodology for resolving undetected mutants to flaws.

  \item\label{goal:performance} {\em \textbf{Minimize overhead.}} We expect \tool to be used by security researchers as well as tool users and developers.
Hence, we must ensure that \tool is efficient so as to promote a wide-scale deployment and community-based use of \tool.
\end{enumerate}

\myparagraph{Threat Model} The design goals delineated above are ultimately meant to address a specific threats that arise from static analyses that are \textit{thought} to be sound -- but are not actually not sound -- leading to undiscovered security vulnerabilities in Android apps.
{\tool} is designed to help security researchers evaluate tools that  detect vulnerabilities (\eg{} SSL misuse), {\em and more importantly}, tools that detect malicious or suspicious behavior (\eg{} data leaks). Thus, the security operators and mutation schemes defined in this paper are of an adversarial nature. That is, behavior like ``data leaks'' is intentionally malicious/curious, and generally not attributed to accidental vulnerabilities. Therefore, to evaluate the soundness of existing tools that detect such behavior, \tool has to develop mutants that mimic such adversarial behavior as well, by defining mutation schemes of an adversarial nature.  This is the key difference between \tool and prior work on fault/vulnerability injection (e.g., LAVA~\cite{dhk+16}) that assumes the mutated program to be benign.

\section{Design}
\label{sec:design}
%\vspace{-0.2em}
\begin{figure}[!tbp]
    \centering
    \begin{minipage}[b]{0.47\textwidth}
      \includegraphics[width=\textwidth]{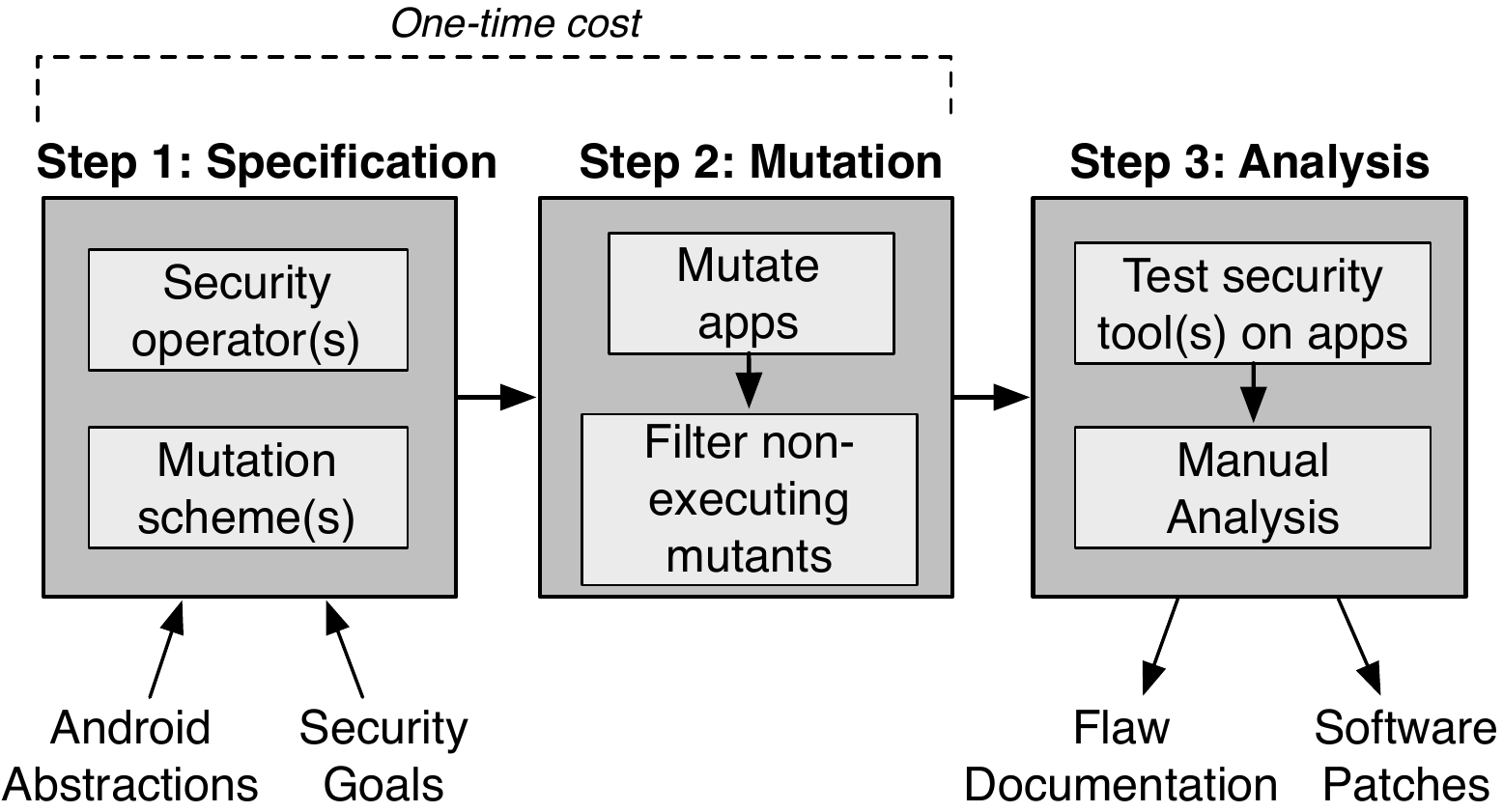}
      \caption{\small The components and process of the \tool{}.}\label{fig:design}
    \end{minipage}
    \hfill
    \begin{minipage}[b]{0.45\textwidth}
      \includegraphics[width=\textwidth]{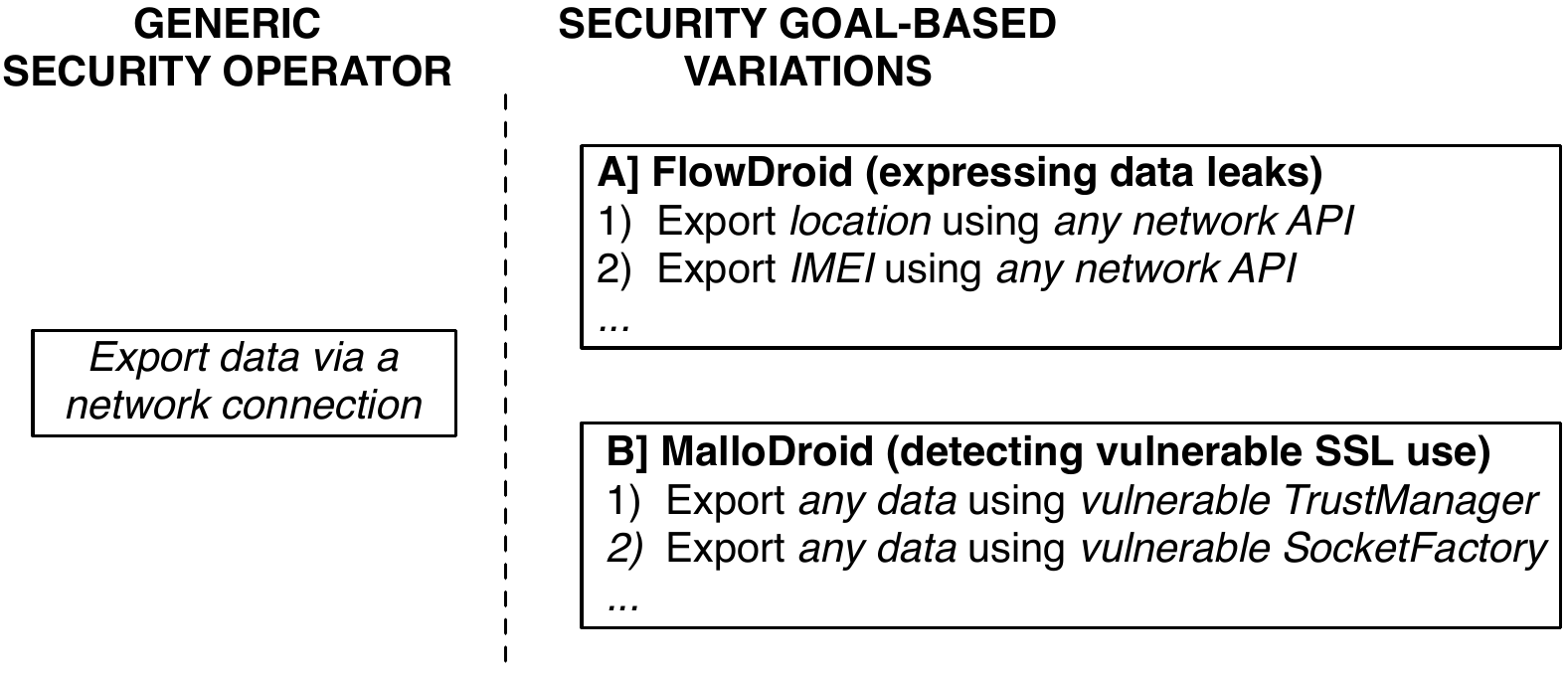}
      \caption{{\small A generic ``network export'' security operator, and its
             more fine-grained instantiations in the context of
             FlowDroid~\cite{arf+14} and
             MalloDroid~\cite{fhm+12}.}}\label{fig:instantiation}
    \end{minipage}
    \vspace{-1em}
\end{figure}

Figure~\ref{fig:design} provides a conceptual description of the process  followed by \tool, which consists of three main steps. In Step 1, we {\em specify} the security operators and mutation schemes that are relevant to the security goals of the tool being evaluated (\eg{} data leak detection), as well as certain unique abstractions of Android that separately motivate this analysis. In Step 2, we {\em mutate} one or more Android apps using the security operators and defined mutation schemes using a {\em Mutation Engine (ME)}. After this step each app is said to contain one or more mutants. To maximize effectiveness, mutation schemes in \tool{} stipulate that mutants should be systematically injected into all potential locations in code where operators can be instantiated. In order to limit the effort required for manual analysis due to potentially large numbers of mutants, we first filter out the non-executing mutants in the Android app(s) using a dynamic {\em Execution Engine (EE)} (Section~\ref{sec:implementation}). In Step 3, we apply the security tool under investigation to {\em analyze} the mutated app, leading it to detect some or all of the mutants as anomalies. We perform a methodological manual analysis of the undetected mutants, which may lead to documentation of flaws, and software patches.
Note that tools sharing a security goal (\eg{} FlowDroid~\cite{arf+14},  Argus~\cite{wror14}, HornDroid~\cite{cgm16} and BlueSeal~\cite{svt+14} all detect data leaks) can be analyzed using the same security operators and mutation schemes, and hence the mutated apps, significantly reducing the overall cost of operating \tool{} (Goal~\ref{goal:performance}).

This section describes the design of \tool, including the additional contributions made in this extension of our \usenix study~\cite{richie18}.
Moreover, in Section~\ref{sec:scheme-generality}, we describe a general approach for leveraging the abstractions introduced in \tool to evaluate security tools built for goals other than data leak detection (\eg, cryptographic API misuse detection).

\subsection{Security Operators}
\label{sec:operators}

A security operator is a description of the unwanted behavior that the security  tool being analyzed aims to detect. When designing security operators, we are faced with an important question: {\em what do we want to express?} Specifically, the operator might be too coarse or fine-grained; finding the correct granularity is the key.
For instance, defining operators specific to the implementations of individual  tools may not be scalable. On the contrary, defining a generic security operator for all the tools may be too simplistic to be effective. Consider the following example:\\ Figure~\ref{fig:instantiation} describes the limitation of using a generic security operator that describes code which ``exports data to the network''. Depending on the tool being evaluated, we may need a unique, fine-grained, specification of this operator. For example, for evaluating FlowDroid~\cite{arf+14}, we may need to express the specific types of private data that can be exported via any of the network APIs, \ie{} the data portion of the operator is more important than what network API is used. However, for evaluating a tool that detects vulnerable SSL connections (\eg{} CryptoLint~\cite{ebfk13}), we may want to express the use of vulnerable SSL APIs (\ie{} of SSL classes that can be overridden, such as a custom TrustManager that trusts all certificates) without much concern for what data is exported. That is, the requirements are practically orthogonal for these two use cases, rendering a generic operator useless, whereas precisely designing tool-specific operators may not scale.

In \tool, we take a balanced approach to solve this problem: instead of tying a  security operator to a specific tool, we define it in terms of the {\em security goal} of the concerned tool (Goal~\ref{goal:operator}).
Because the security goal influences the properties exhibited by a security analysis, security operators designed with a particular goal in consideration would apply to all the tools that claim to have that security goal, hence making them feasible and scalable to design.
For instance, a security operator that reads information from a private source (\eg{} IMEI, location) and exports it to a public sink (\eg{} the device log, storage) would be appropriate to use for all the tools that claim to detect private data leaks (\eg{} Argus~\cite{wror14}, HornDroid~\cite{cgm16}, BlueSeal~\cite{svt+14}) (\eg, see Listing~\ref{lst:operator_leak} for one such implemented operator).
%For instance, one of our implemented operators for evaluating tools that detect data leaks is as described in Listing~\ref{lst:operator_leak}.
Moreover, security operators generalize to other security goals as well; a simple operator for evaluating tools that detect vulnerable SSL use (\eg{} MalloDroid) could add a \verb|TrustManager| with a vulnerable \verb|isServerTrusted| method that returns true.

\begin{lstlisting}[basicstyle=\ttfamily\scriptsize,float,caption={{\small
    Security operator that injects a data leak from the Calendar API
    access to the device log.}},belowcaptionskip=-8mm,label=lst:operator_leak,emph={Inject},emphstyle=\bfseries]
  Inject: 
    String dataLeak## = java.util.Calendar.getInstance().getTimeZone().getDisplayName();
    android.util.Log.d("leak-##", dataLeak##);
  \end{lstlisting}

To derive security operators at the granularity of the security goal, we must  examine the claims made by existing tools; \ie{} security tools must certainly detect the unwanted behavior that they claim to detect, unless affected by some unsound design choice that hinders detection.  We inspect the following sources to precisely identify the security flaws considered by tools:

\myparagraph{1) Research Papers} The tool's research paper is often the primary  source of information about what unwanted behavior a tool seeks to detect. We inspect the properties and variations of the unwanted behavior as described in the paper, as well as the examples provided, to formulate security operator specifications for injecting the unwanted behavior in an app. However, we do not create operators using the limitations and assumptions already documented in the paper or well-known in general (\eg{} reflection or dynamic code loading), as \tool{} seeks to find unknown assumptions.

\myparagraph{2) Open source tool documentation} Due to space limitations or tool  evolution over time, research papers may not have the most complete or up-to-date information considering what security flaws a tool can actually address. We used tool documentation available in online appendices and open source repositories to fill this knowledge gap.

\myparagraph{3) Testing toolkits} Manually-curated testing toolkits (\eg{}  DroidBench~\cite{arf+14}) may be available and may provide examples of baseline operators.

\subsection{Mutation Schemes}
\label{sec:mutation-scheme}
To enable the security evaluation of static analysis tools, \tool{} must seed  mutations within Android apps.
For this purpose, we define \textit{mutation schemes}, \ie, the methods for choosing \textit{where} to apply security operators to inject mutations within Android apps.

Our design of mutation schemes leverages a number of factors:
{\sf (1)} Android's unique  abstractions, {\sf (2)}, the intent to over-approximate reachability for coverage, and {\sf (3)} the security goal of the tool being analyzed.
% \add{
  We design mutation scheme \textit{strategies} based on these factors (Section~\ref{sec:android-scheme}$\rightarrow$Section~\ref{sec:scheme_security_goal}), and describe them in the context of our running example first described in Section~\ref{sec:motivation} (but elaborated as follows):
% }

Recall that FlowDroid~\cite{arf+14}, the target of our analysis in  Section~\ref{sec:motivation}, detects data leaks in Android apps. Hence, FlowDroid loosely defines a data leak as a flow from a sensitive \textit{source} of information to some \textit{sink} that exports it. FlowDroid lists all of the sources and sinks within a configurable ``SourcesAndSinks.txt'' file in its tool documentation.
A simple data leak mutation may be implemented by using the data leak operator described previously, with a source (\eg, {\sf \small java.util.Calendar.getTimeZone()}) and sink (\eg, {\sf \small android.util.Log.d()}) from this file.

The strategies described in the rest of this section guide our implementation of {\em four} specific mutation schemes, using which we seed this data leak across target Android applications (see Section~\ref{sec:implementation} for implementation).
In particular, we significantly enhance the goal-based mutation scheme strategy by introducing scope as a factor, and implement a scope-based mutation-scheme (Section~\ref{sec:implementation}), which further increases the expressiveness of the mutation introduced by \tool.

\subsubsection{\textbf{Mutation Scheme Strategy 1:} Leveraging Android Abstractions}
\label{sec:android-scheme}
The Android platform and app model support numerous abstractions that pose  challenges to static analysis. One commonly stated example is the absence of a {\sf Main} method as an entry-point into the app, which compels static analysis tools to scan for the various entry points, and treat them all similarly to a traditional {\sf Main} method~\cite{arf+14, Halavanalli:2013}.

Based on our domain knowledge of Android and its security, we choose the  following features as a starting point in a mutation scheme strategy that models unique aspects of Android, and  more importantly, tests the ability of analysis tools to detect unwanted behavior placed within these features (Goal~\ref{goal:android}):

\myparagraph{1) Activity and Fragment lifecycle}  Android apps are organized  into a number of {\em activity} components, which form the user interface (UI) of the app. The activity lifecycle is controlled via a set of callbacks, which are executed whenever an app is launched, paused, closed, started, or stopped~\cite{android-lifecycle}. In addition, {\sf Fragments} are UI elements that possess similar callbacks, although they are often used in a manner secondary to activities.  We design our mutation scheme to place mutants within methods of fragments and activities where applicable, so as to validate a tool's ability to model the activity and fragment life-cycles.

\myparagraph{2) Callbacks} Because much of Android relies on callbacks triggered by events, these callbacks  pose a significant challenge to traditional static analyses, as their code can be executed asynchronously in several different potential orders. We place mutants within these asynchronous callbacks to validate the tools' ability to soundly model the asynchronous nature of Android.  For instance, consider the example in Listing~\ref{lst:onClick},
where the {\sf onClick()} callback can execute at any point of time.

\begin{lstlisting}[basicstyle=\ttfamily\scriptsize,float,caption={{\small
    Dynamically created {\sf onClick} callback}},
    belowcaptionskip=-7mm,label=lst:onClick,
    emph={},emphstyle=\bfseries]
  final Button button = findViewById(R.id.button_id);
  button.setOnClickListener(new View.OnClickListener() {
      public void onClick(View v) {
          // Code here executes on main thread after user presses button
          }
      });
  \end{lstlisting}

\myparagraph{3) Intent messages} Android apps communicate with one another and  listen for system-level events using Intents, Intent Filters, and Broadcast Receivers~\cite{android-intents,android-broadcasts}. Specifically, Intent Filters and Broadcast Receivers form another major set of callbacks into the app. Moreover, Broadcast Receivers can be dynamically registered. Our mutation scheme not only places mutants in the statically registered callbacks such as those triggered by Intent Filters in the app's Android Manifest, but also callbacks dynamically registered within the program, and even within other callbacks, \ie{} recursively. For instance, we generate a dynamically registered broadcast receiver inside another dynamically registered broadcast receiver, and instantiate the security operator within the inner broadcast receiver (see Listing~\ref{lst:inception} in the Appendix for the code).

\myparagraph{4) XML resource files} Although Android apps are primarily written  in Java, they additionally include resource files that establish callbacks. Such resource files also allow the developer to register for callbacks from an action on a UI object (\eg{} the {\sf onClick} event, for callbacks on a button being touched). As described previously, static analysis tools often list these callbacks on par with the {\sf Main} function, \ie{} as one of the many entry points into the app. We incorporate these resource files into our mutation scheme, \ie{} mutate them to call our specific callback methods.

%\vspace{-0.3em}
\subsubsection{\textbf{Mutation Scheme Strategy 2:} Evaluating Reachability}
\label{sec:reachability-scheme}
%\vspace{-0.2em}

The objective behind this simple, but important, mutation scheme is to  exercise the reachability analysis of the tool being evaluated. We inject mutants (\eg{} data leaks from our example) at the start of every method in the app.
% \add{
Furthermore, as part of this extended study, we add leaks at the class declaration-level as well.
% }
While the previous schemes add methods to the app (\eg{} new callbacks), this scheme simply verifies if the app successfully models the bare minimum.

%\vspace{-0.3em}
\subsubsection{\textbf{Mutation Scheme Strategy 3:}
\label{sec:scheme_security_goal}
Leveraging the Security Goal}\label{sec:goal-scheme}
%\vspace{-0.2em}

Like security operators, mutation schemes may also be designed in a way that  accounts for the security goal of the tool being evaluated (Goal~\ref{goal:operator}). Such schemes may be applied to any tool with a similar objective. In keeping with our motivating example (Section~\ref{sec:motivation}) and our evaluation (Section~\ref{sec:eval}), we develop an example mutation scheme strategy that can be specifically applied to evaluate data leak detectors. This strategy can be instantiated in terms of the following three methods of seeding mutants:

\begin{lstlisting}[style=toplisting,basicstyle=\ttfamily\scriptsize,caption={{\small
    Complex Path Operator Placement}},belowcaptionskip=-5mm,label=lst:complex,emph={Inject},emphstyle=\bfseries]
String dataLeak0 = java.util.Calendar.getInstance().getTimeZone().getDisplayName();
String[] leakArRay0 = new String[] {"n/a", dataLeak0};
String dataLeakPath0 = leakArRay0[leakArRay0.length - 1];
android.util.Log.d("leak-0", dataLeakPath0);
\end{lstlisting}
\begin{lstlisting}[style=toplisting,basicstyle=\ttfamily\scriptsize,caption={{\small
    Scope based operator placement at different levels of inheritance.}},belowcaptionskip=-5mm,label=lst:scope,emph={Inject},emphstyle=\bfseries]
public class ParentClass {
    String dataLeak = "";
    int methodA(){
        android.util.Log.d("leak-0-1", dataLeak);
        return 1; }
    class ChildClass{
        int childMethodA(){
            dataLeak = java.util.Calendar.getInstance().getTimeZone().getDisplayName();
            android.util.Log.d("leak-0-0", dataLeak);
            return 1; }}}
 \end{lstlisting}

\myparagraph{1) Taint-based operator placement} This placement methodology tests the tools' ability to recognize an asynchronous  ordering of callbacks, by placing {\em the source in one callback and the sink in another}. The execution of the source and sink may be triggered due to the user, and the app developer (\ie{} especially a malicious adversary) may craft the mutation scheme specifically so that the sources and sinks lie on callbacks that generally execute in sequence. However, this sequence may not be observable through just static analysis. A simple example is collecting the source data in the {\sf onStart()} callback, and leaking it in the {\sf onResume()} callback.
As per the activity lifecycle, the {\sf onResume()} callback {\em always}  executes right after the {\sf onStart()} callback.

\myparagraph{2) Complex-Path operator placement} Our preliminary analysis demonstrated that static analysis tools may sometimes  stop after an arbitrary number of hops when analyzing a call graph, for performance reasons. This finding motivated the complex-path operator placement. In this scheme, we make the path between source and sink as complex as possible (\ie{} which is ordinarily one line of code, as seen in Listing~\ref{lst:operator_leak}). That is, the design of this scheme allows the injection of code along the path from source to sink based on a set of predefined rules. For our evaluation, we instantiate this scheme with a rule that recreates the String variable saved by the source, by passing each character of the string into a {\sf StringBuilder}, then sending the resulting string to the sink as shown in Listing~\ref{lst:complex}. \tool{} allows the analyst to dynamically implement such rules, as long as the input and output are both strings, and the rule complicates the path between them by sending the input through an arbitrary set of transformations.
% \vspace{1em}

\myparagraph{3) Scope-based operator placement}
% \add{
This scope-based placement methodology was not defined in our \usenix paper~\cite{richie18}, and is a new addition to \tool as part of this extension. As the name suggests, \tool can inject code by analyzing scopes based on \textit{visibility}.
For example, as shown in Listing~\ref{lst:scope}, {\sf childMethodA} is visible from both {\sf ChildClass} and {\sf ParentClass}.
Therefore, we declare a variable {\sf dataLeak} at the {\sf ParentClass}, while assigning the leak source at the  {\sf childMethodA}.
Consequently, we insert corresponding sinks in {\sf childMethodA} and {\sf methodA}.
Note that this scheme is not restricted by callbacks and can be useful for different variations of method declarations.
As application developers may arbitrarily organize class scopes, seeding mutants into applications using this scheme generally results in interesting and often complicated mutant placement, which further assists in stress-testing security techniques that assume an adversarial threat model (e.g., data leak detectors).
% }

In a traditional mutation analysis setting, the mutation placement strategy  would seek to minimize the number of non-compilable mutants.
However, as our goal is to evaluate the soundness of Android security tools, we design our mutation scheme to over-approximate.
Once the mutated apps are created, for a feasible analysis, we pass them through a dynamic filter that removes the mutants that cannot be executed, ensuring that the mutants that each security tool is evaluated against are all executable \ie, the data leaks can indeed happen in runtime.

% \add{
Note that although mutation schemes using the first two strategies
(Section~\ref{sec:android-scheme} and Section~\ref{sec:reachability-scheme}) may be generally applied to any type of static analysis tool (\eg{} SSL vulnerability and malware detectors), the third strategy, as the description suggests, will need to be adapted per security goal (\eg, data leak detection).
 We elaborate on this point by presenting a general approach for leveraging our mutation abstractions for other security goals, in Section~\ref{sec:scheme-generality}.
% }

% \vspace{-0.2em}
\subsection{Analysis Feasibility \& Methodology}
\label{sec:feasibility}
% \vspace{-0.3em}
\begin{figure}[t]
  \centering
  \includegraphics[width=1.4in]{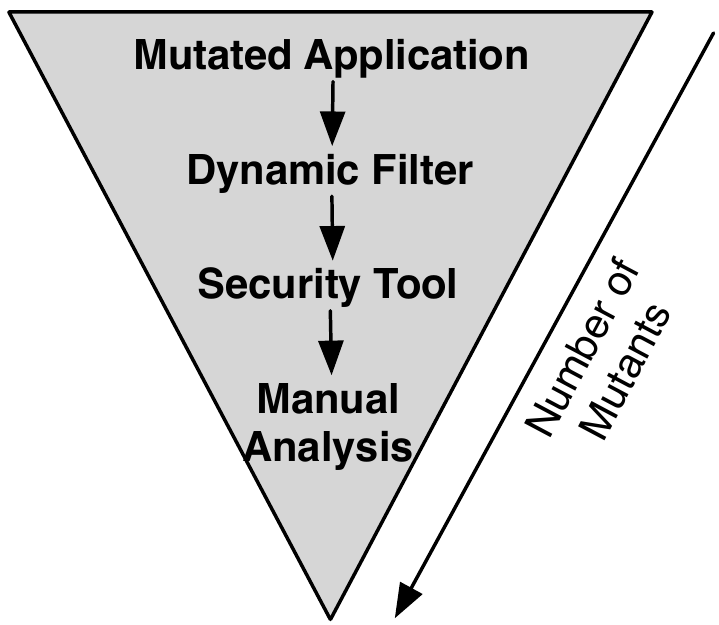}
  \vspace{-1em}
  \Description{A downward triangle that shows that number of mtuatants reduce
  through each stage} \caption{{\small The number of mutants (\eg{} data leaks)
  to analyze drastically reduces at every stage in the process.}}\label{fig:filter}
\vspace{-2em}
\end{figure}

% \AMIT{Changes here}
\tool{} reduces manual effort by filtering out mutants whose security flaws are not verified by dynamic analysis (Goal~\ref{goal:feasibility}).
As described in Figure~\ref{fig:design}, for any given mutated app, we use a dynamic filter (\ie{} the Execution Engine (EE), described in Section~\ref{sec:implementation}) to purge non-executable leaks.
If a mutant (\eg{} a data leak) exists in the mutated app, but is not confirmed as executable by the filter, (\ie, data does not leak from source to sink), we discard it.
For example, data leaks injected in dead code are filtered out.
Thus, when the Android security tools are applied to the mutated apps, only mutants that were executed by EE are considered.

Furthermore, after the security tools were applied to mutant apps, only  \textit{undetected} mutants are considered during analyst analysis. The reduction in the number of mutants subject to analysis at each step of the \tool{} process is illustrated in Figure~\ref{fig:filter}.

The following methodology is used by an analyst for each undetected mutant after testing a given security tool to isolate and confirm flaws:

\myparagraph{1) Identifying the Source and Sink} During mutant generation, \tools{} ME injects a unique mutant identifier, as well as the source and sink using {\sf util.Log.d} statements. Thus, for each undetected mutant, an analyst simply looks up the unique IDs in the source to derive the source and sink.

	\myparagraph{2) Performing Leak Call-Chain Analysis} Since the data leaks under analysis went undetected by a given static analysis tool, this implies that there exists one (or multiple) method call sequences (\ie, call-chains) invoking the source and sink that could not be modeled by the tool.  Thus, a security analyst inspects the code of a mutated app and identifies the observable call sequences from various entry points.  This is aided by dynamic information from the EE so that an analyst can examine the order of execution of detected data leaks to infer the propagation of leaks through different call chains.

\myparagraph{3) Synthesizing Minimal Examples} For each of the identified call sequences invoking a given undetected data leak's source and sink, an analyst then attempts to synthesize a minimal example by re-creating the call sequence using only the required Android APIs or method calls from the mutated app. This info is then inserted into a pre-defined skeleton app project so that it can be again analyzed by the security tools to confirm a flaw.

\myparagraph{4) Validating the Minimal Example} Once the minimal example has been synthesized by the analyst, it must be validated against the security tool that failed to detect it earlier.  If the tool fails to detect the minimal example, then the process ends with the confirmation of a flaw in the tool. If the tool is able to detect the examples, the analyst can either iteratively refine the examples, or discard the mutant, and move on to the next example.
% \add{
\subsection{\textbf{Leveraging Security Operators and Mutation Schemes for other Security Goals}}
\label{sec:scheme-generality}
Our design describes the \tool framework in terms of the security goal of data leak detection, primarily due to the popularity of the goal, as well as the proliferation of data leak detectors in academic research.
However, \tool may be applied as a general framework, to evaluate detection techniques that target other security goals.
We briefly describe the generality of \tool, and specifically, its abstractions of security operators and mutation schemes for evaluating tools that detect {\em cryptographic-API misuse}, which is the {\em second most} prolific cause of vulnerabilities, after data leaks~\cite{veracode-report}.
In this context, \tool will evaluate tools such as CryptoGuard~\cite{rxa+19} and CrySL~\cite{knr+17}, which try to detect misuse of crypto APIs.

\tools abstractions may be naturally leveraged to evaluate crypto-API misuse detectors.
To elaborate, the {\em security operators} for evaluating such tools would represent well-known cryptographic vulnerabilities, such as passing a weak algorithm name as parameter to an encryption related cryptography API.
For example, one operator could be represented by passing insecure parameters in an API such as \texttt{Cipher.getInstance()}, such as {\em only passing {\tt AES}} as a parameter, which would initialize an insecure default cipher that use the {\tt ECB} mode.

Similarly, \tools mutation schemes may also be directly leveraged to place such mutations in a hard-to-detect manner.
First, the scheme that evaluates reachability (Section~\ref{sec:reachability-scheme}) can be directly applied, by placing such vulnerable API invocations (i.e., mutations) at as many reachable locations within the target app as possible.
Similarly, the Android-specific scheme-strategies (Section~\ref{sec:android-scheme}) can be leveraged to place the vulnerable invocation in commonly used Android abstractions such as fragments and broadcast receivers.
Finally, the security goal-based scheme-strategies (Section~\ref{sec:scheme_security_goal}) can be leveraged to evaluate crypto-API misuse detectors as well.
For example, we can consider the parameter passed to the \texttt{Cipher.getInstance()} API, \ie, the {\tt AES} string, as a source value, and the API call as the sink, and use the taint-based mutation scheme to distribute these source and sink values across different Android lifecycle methods, such that they would execute in the right sequence, but be hard to detect.
Similarly, our newly defined scope-based mutation scheme can be used to inject the source parameter and the API in different program scopes, in a way that would execute flawlessly at runtime, but would be hard to detect.
Other than syntactical changes required to implement the security operator
% \add{
  (\ie, defining new \textit{source-sink pairs}, but with Cipher APIs)
% }
and adjusting it for mutation using the schemes
% \add{
(\ie, scoping them within required \textit{try-catch} block)
% }
, no other changes will be required to the framework, for making it applicable for evaluating crypto-API misuse detectors. We elaborate on the estimated effort required to make these changes in Section~\ref{sec:limitations}.
% }

\section{Implementation}
\label{sec:implementation}

This section provides the implementation details of \tool{}'s components: {\sf (1)} the ME for  mutating apps, and {\sf (2)} the EE for exercising and filtering out non-executable mutants. We have made \tool{} available to the research community~\cite{appendix}, along with all the data and code generated or used.

\myparagraph{1. Mutation Engine (ME)} The ME allows \tool{} to automatically  mutate apps according to a fixed set of security operators and mutation schemes.
The ME is implemented in Java and extends the \mdroid{} mutation framework for Android~\cite{lbt+17}.
% \add{
More specifically, \tools ME implements the seeding of mutants according to our defined \textit{mutation schemes}. This required extensions to \mdroid's implemented static analyses to identify a more diverse array of source code locations
(\eg, analyzing the visibility scope for the {\em scope-based operator placement} scheme).
\mdroid's previous implementation only supported identifying strings and specific API patterns. These additions help to make \tools ME generic, as it could be applied to support additional security operators or mutation schemes in the future.  To achieve these extensions, we designed ME to do the following:
% }

Firstly, the ME derives a mutant injection profile (MIP) of all possible injection points for a given mutation scheme, security operator, and target app source code.
The MIP is derived through one of two types of analysis: (i) text-based parsing and matching of \texttt{xml} files in the case of app resources; or (ii) using Abstract Syntax Tree (AST) based analysis for identifying potential injection points in code.
\tool{} takes a systematic approach toward applying mutants to a target app, and for each mutant location stipulated by the MIP for a given app, a mutant is seeded.
The injection process additionally uses either text or AST-based code transformation rules to modify the code or resource files.
In the context of our evaluation, \tool{} further marks injected mutants in the source code with log-based indicators that include a unique identifier for each mutant, as well as the source and sink for the injected leak.
This information can be customized for future security operators and exported as a ledger that tracks mutant data.
\tool{} can be extended to additional security operators and mutation schemes by adding methods to derive the MIP and perform target code transformations.

Given the time cost in running the studied security-focused static analysis  tools on a set of \texttt{apks}, \tool{} breaks from the process used by traditional mutation analysis frameworks that seed each mutant into a separate program version, and seeds all mutants into a single version of a target app. Finally, the target app is automatically compiled using its build system (\eg, Gradle~\cite{gradle}, ant~\cite{ant}) so that it can be dynamically analyzed by the EE.

\myparagraph{2. Mutation Schemes}
We implemented four mutation schemes using the strategies outlined in Section~\ref{sec:mutation-scheme}.
First, we implemented the \reachability{} scheme, using the reachability-based strategy (Section~\ref{sec:reachability-scheme}).
Then, we implemented three other schemes, \scope{}, \taint{}, and \complexreachability{}, using the security goal-based strategy (Section~\ref{sec:goal-scheme}).
Finally, we integrated the Android-specific mutation scheme strategy (Section~\ref{sec:android-scheme}) in these four schemes by prioritizing placement into several abstractions unique to Android (\eg, fragments, receivers, asynchronous task classes).
Among these, the \scope{} based scheme was not implemented in our prior \usenix paper~\cite{richie18}.

\myparagraph{3. Execution Engine (EE)}  To facilitate a feasible manual analysis  of the mutants that are undetected by a security analysis tool, \tool{} uses the EE to dynamically analyze target apps, verifying whether or not injected mutants can be executed in practice.
% \add {
To further elaborate with respect to our implementation, an \textit{executable mutant} represents a data leak from source to sink as observed through our execution engine.
% }
This EE builds upon prior work in automated input generation for Android apps by adapting the systematic exploration strategies from the {\sc \small CrashScope} tool~\cite{Moran:ICST16, Moran:ICSE17} to explore a target app's GUI.
% \add{
We made several improvements to {\sc \small CrashScope} in order to tailor the automated app execution to the goal of discovering as many {\em executable} mutants as possible.
For instance, {\sc \small CrashScope}'s execution strategies were originally intended to uncover crashes in Android apps and thus included several mechanisms that executed common crash inducing actions on apps including rotating the screen, and injecting text with special characters.
In order to be used in \tool, we discarded these crash inducing operations, and developed strategies that focused upon uncovering as many execution states of the app as possible.
% \add{
Additionally, as part of extending our \usenix paper~\cite{richie18}, we made a practical improvement to the manner in which {\sc \small CrashScope} analyzes applications as part of our current study.
% }
That is, during initial testing we discovered that even after uninstalling an application from an Android virtual device used in {\sc \small CrashScope}, certain background services may persist, which can contaminate the runtime mutant execution logs across applications.
We addressed this problem by modifying {\sc \small CrashScope} to instantiate a new, clean Android virtual device for each application analyzed, thereby fully isolating application execution logs.
% }

We discuss the limitations of the EE in Section~\ref{sec:limitations} (see the Appendix for additional details).

\section{Evaluation Overview}
\label{sec:eval}

% \add{
The primary objective of this evaluation is to measure  the effectiveness of \tool{} at uncovering flaws in security-focused static analysis tools for Android apps, and to demonstrate the extent of such flaws.
As a case study, we focus our evaluation on the security goal of data leak detection, which has received significant attention from the security research community in the last decade~\cite{egc+10, ekkv11, gcec12, lbk+14, ekkv11, lbb+15,jsg+08,Rasthofer:2014,Li:2014}.
% \add{
Our evaluation is guided by 6 key research questions that provide an objective measure of the effectiveness and practicality of \tool, while also shedding light on the general nature of unsound decisions/flaws in static analysis security tools:
% }
\begin{enumerate}[label=\textbf{RQ$_\arabic*$},ref=\textbf{RQ$_\arabic*$}]

    \item\label{rq:findings} {\bf \em Effectiveness of \tool.} {\em Can \tool find security problems in static analysis tools for Android, and help resolve them to flaws/unsound choices?}%
	% \add{
	\item\label{rq:effectiveness} {\em {\bf \em Relative effectiveness of mutation schemes.\footnote{
		\ref{rq:effectiveness} was not investigated in our \usenix paper~\cite{richie18}.}} How effective are the individual mutation schemes for (a) generating executable mutants and (b) discovering security flaws?}
	% }
	\item\label{rq:parent_inherit} {\em {\bf \em Propagation of flaws.} Are flaws inherited when a tool is reused, or built based on similar principles?}
	\item\label{rq:patch} {\em {\bf \em Addressing flaws.} Are all flaws unearthed by \tool hard to address, or can some be patched?}
	\item\label{rq:feasible} {\em {\bf \em Scalability of \tool.} Does the semi-automated methodology of \tool for analyzing mutants allow for a feasible analysis (\ie, in terms of the manual effort)?}
	\item\label{rq:perf} {\em {\bf \em Performance of \tool.} What is the runtime performance of \tool?}

\end{enumerate}
%}

% \add{
To address \ref{rq:findings}$\rightarrow$\ref{rq:perf}, we performed several experiments for a period of over two years (\ie, October 2017 $\rightarrow$  January 2020).
%Section 7: Executing mSE
We started by creating a data-leak security operator (\ie, as described in Section~\ref{sec:operators}), and used \tools expressive mutation schemes (Section~\ref{sec:mutation-scheme}) to seed the corresponding mutated code in a set of $15$ open source Android applications obtained from F-droid~\cite{fdroid} (see Table~\ref{tbl:app_list} in the Appendix for the list), creating \countinsertedtotalleaks mutants representing injected data leaks.
Section~\ref{sec:executing_tools} describes this process, along with the refinement made possibly by \tools EE, and addresses questions pertaining to \tools intrinsic evaluation (\ref{rq:effectiveness},~\ref{rq:feasible} and~\ref{rq:perf}).
We selected three prominent data leak detectors, namely, FlowDroid~\cite{arf+14}, Argus~\cite{wror14} (previously known as AmanDroid), and HornDroid~\cite{cgm16}, as the target of our in-depth evaluation using the end-to-end approach as described in Section~\ref{sec:design}.
% \add{
The in-depth study of FlowDroid was reported in our \usenix{} paper~\cite{richie18} and is not re-performed in this study.
% }
Due to the longitudinal nature of this study, we strived to use the latest release of the data leak detectors whenever available.
% \add{
Section~\ref{sec:indepth_eval} describes this evaluation, the $13$ flaws that we previously discovered (\ref{rq:findings}) in FlowDroid and the $12$ new flaws found in HornDroid and Argus.
Further, we also briefly describe the one flaw that we could fix (\ref{rq:patch}), and the relative effectiveness of \tools mutation schemes in unearthing flaws (\ref{rq:effectiveness}).
% }
We developed minimal examples of the discovered flaws, and performed a {\em flaw propagation study} (Section~\ref{sec:propagation_study_pre}) to discover the extent to which flaws discovered in one tool manifest in others developed for the same security goal (\ref{rq:parent_inherit}).
Particularly, we used four additional data leak detection tools, \ie, in addition to the three tools analyzed in depth in Section~\ref{sec:indepth_eval}, bringing the total to seven tools, for which propagation was studied.
Note that compared to our \usenix paper~\cite{richie18}, which only studied the propagation of flaws discovered in FlowDroid, we describe the propagation of flaws discovered in three tools, namely FlowDroid, Argus, and HornDroid.
Our results demonstrate that flaws generally propagate to other tools, and more so if the tools rely on common design principles.
% }

% \input{texfiles/mutation-engine-study.tex}

% Running Muse on static analysis tools
\section{Executing \tool to Create Mutants Representing Data Leaks}
\label{sec:executing_tools}

We applied \tool to $15$ target Android apps obtained from F-Droid~\cite{fdroid}, and created \countinsertedtotalleaks{} mutants (\ie, data leaks).\footnote{We use ``mutants'' and ``leaks'' interchangeably to refer to data-leak-related mutants used in the evaluation, \ie, Sections~\ref{sec:eval}$\rightarrow$\ref{sec:propagation_study_pre}.}
These leaks were generated by the \tools Mutation Engine (ME) using the data-leak security operator,
% \add{
and the four mutation schemes described in Section~\ref{sec:implementation}, namely, {\sf (1)} {\em reachability}, {\sf (2)} {\em scope}, {\sf (3)} {\em taint}, and {\sf (4)} {\em complex reachability}.
% }

\begin{table}[t]
    \caption{
        The number of leaks inserted by \tool, and the final number marked as executable by \tools EE{}. Note that the ``-'' indicates that the scheme is not applicable to a particular app, due to the app's particular characteristics (\eg, the absence of fragments)
        }

    \vspace{-0.5em}
    \scriptsize
    \centering

    \begin{tabular*}{\textwidth}{l @{\extracolsep{\fill}} l l l l l l l l}
        \toprule
        \multirow{2}[2]{*}{\textbf{App ID}} & \multicolumn{4}{c}{\textbf{Inserted Leaks per Scheme}} & \multicolumn{4}{c}{\textbf{Executable Leaks per Scheme}}    \\
        \cmidrule(){2-5}\cmidrule{6-9}
& \textbf{Reachability} & \textbf{Scope} & \textbf{Taint} & \textbf{\begin{tabular}[c]{@{}l@{}}Complex \\ Reachability\end{tabular}} & \textbf{Reachability} & \textbf{Scope} & \textbf{Taint} & \textbf{\begin{tabular}[c]{@{}l@{}}Complex\\ Reachability\end{tabular}} \\
        % \cmidrule(r){1-1}\cmidrule(lr){2-5}\cmidrule(lr){6-9}
        \midrule
        App 01  & 24   & 48   & 66    & 22   & 13  ($\approx$54\%)  & 12  ($\approx$12\%)   & 8    ($\approx$12\%) & 11   ($\approx$50\%)\\
        App 02  & 106  & -    & 231   & 83   & 63  ($\approx$59\%)  & -                     & 80   ($\approx$34\%) & 44   ($\approx$14\%)\\
        App 03  & 248  & -    & 668   & 191  & 51  ($\approx$20\%)  & -                     & 162  ($\approx$24\%) & 36   ($\approx$26\%)\\
        App 04  & 45   & 192  & 346   & 41   & 20  ($\approx$44\%)  & 156 ($\approx$81\%)   & 51   ($\approx$14\%) & 23   ($\approx$2\%)\\
        App 05  & 3181 & -    & 12104 & 2777 & 598 ($\approx$18\%)  & -                     & 1362 ($\approx$11\%) & 508  ($\approx$42\%)\\
        App 06  & 434  & -    & 2699  & 384  & 57  ($\approx$13\%)  & -                     & 110  ($\approx$4\%)  & 51   ($\approx$5\%)\\
        App 07  & 204  & -    & 551   & 174  & 111 ($\approx$54\%)  & -                     & 243  ($\approx$44\%) & 95   ($\approx$34\%)\\
        App 08  & 975  & 1193 & 5266  & 839  & 215 ($\approx$22\%)  & 56  ($\approx$4\%)    & 567  ($\approx$10\%) & 168  ($\approx$32\%)\\
        App 09  & 23   & 8    & 20    & 21   & 15  ($\approx$65\%)  & 6   ($\approx$75\%)   & 13   ($\approx$65\%) & 12   ($\approx$7\%)\\
        App 10  & 476  & -    & 5283  & 449  & 59  ($\approx$12\%)  & -                     & 681  ($\approx$12\%) & 54   ($\approx$11\%)\\
        App 11  & 316  & 1111 & 827   & 277  & 50  ($\approx$15\%)  & 287 ($\approx$25\%)   & 83   ($\approx$10\%) & 41   ($\approx$5\%)\\
        App 12  & 250  & 354  & 428   & 213  & 77  ($\approx$30\%)  & 111 ($\approx$31\%)   & 59   ($\approx$13\%) & 38   ($\approx$7\%)\\
        App 13  & 156  & 203  & 828   & 147  & 89  ($\approx$57\%)  & 112 ($\approx$55\%)   & 295  ($\approx$35\%) & 84   ($\approx$71\%)\\
        App 14  & 125  & 844  & 663   & 107  & 79  ($\approx$63\%)  & 55  ($\approx$6\%)    & 27   ($\approx$4\%)  & 59   ($\approx$9\%)\\
        App 15  & 1304 & 3478 & 2790  & 1143 & 211 ($\approx$16\%)  & 456 ($\approx$13\%)   & 172  ($\approx$6\%)  & 154  ($\approx$41\%)\\
        \midrule
        \textbf{Total}  & 7867 & 7431 & 32770 & 6868 & 1708 ($\approx$22\%) & 1251 ($\approx$16\%) & 3913 ($\approx$11\%) & 1378 ($\approx$20\%) \\
    \bottomrule\label{tbl:leaks_inserted_executable}
    \end{tabular*}
    \vspace{-1.5em}
\end{table}
\myparagraph{Filtering non-executable leaks} We then used our Execution Engine (EE) to filter out non-executable leaks, as described in Section~\ref{sec:implementation},
% \add{
and confirmed \countexecutabletotalleak{} out of \countinsertedtotalleaks{} leaks as executable.
The remaining \countnonexecutabletotalleak{} non-executable leaks were then removed.
% }
Note that this number is independent of the tools involved, \ie, the filtering only happens once, and the mutated APKs can then be passed to any number of tools for analysis.
By filtering out a large number of potentially non-executable leaks
% \add{
(\ie{} {\em \countnonexecutabletotalleak{}/\countinsertedtotalleaks{} or about \countnonexecutabletotalleakpercentage{}\%})
% }
, our dynamic filtering is tremendously effective at reducing the number of mutants used to evaluate security tools, and in turn, the manual effort required to analyze the uncaught mutants, which demonstrates the feasibility of \tool (\ref{rq:feasible}).

% \add{
\myparagraph{Runtime performance} \tool took \counttotalhours{} hours in total to create the mutants and filter out those that were non-executable, which is a one-time cost for each security goal, \ie, which does not have to be repeated for any of the tools we analyze in particular (\ref{rq:perf}).
To elaborate, it took us 74 minutes on average to mutate each of the 15 apps, with minimum and maximum times of 16 and 170 minutes, and a standard deviation of about 53 minutes.
% \add{
The increase in runtime compared to our original work (92 minutes in worst case)~\cite{richie18} is due to two reasons:
(1)  heterogeneity of the applications selected for mutation, and (2)
% }
% Moreover,
a bulk of the time spent can be attributed to our {\em improvements} to CrashScope (see Section~\ref{sec:implementation}) that require the Android virtual device it uses to be recreated more frequently, leading to an increase in runtime.
However, the improvements also increase the reliability of CrashScope's mutant detection by preventing cross-contamination of mutation logs across apps, and hence, are desirable.
% }

% \add{
\myparagraph{Correlation between executable mutants and \tools schemes} To improve our understanding of what factors contribute to more executable leaks, we further examined the number of executable leaks generated as a factor of the mutation schemes used to generate them (\ie, since there was a single security operator used)
% \add{
    in this extended study.
    % }

Table~\ref{tbl:leaks_inserted_executable} shows the number of executable and non-executable leaks seeded using each of the mutation schemes described in Section~\ref{sec:mutation-scheme} (\ref{rq:effectiveness}).
As seen in the table, the \reachability{} and \complexreachability{} result in the insertion of a somewhat similar number of leaks (\ie, 7,867 and 6,868 respectively), and the fraction of leaks deemed executable by our EE is similar as well, \ie, 1,708 (about 22\%) and 1378 (\ie, 20\%) for the \reachability{} and \complexreachability{} scheme, respectively.
Our intuition is that this equivalence results from the inherent similarity in the nature of the two schemes.
Moreover, these two schemes produced the highest fraction of executable mutants (\ie, over 20\%).
Further, the \scope{}-based scheme inserted a total of 7,431 leaks, out of which, 1,251 (\ie, or 16\%) were confirmed as executable by the EE.
Note that the number of leaks inserted using the {\sf Scope}-based scheme for individual apps is highly variable, primarily due to wide variations in the developers' usage of scope.
Finally, the \taint{}-based scheme was the most numerous, both in terms of the number of leaks inserted (\ie, 32,770) as well as the number of executable leaks (\ie, 3,913) it caused.
However, the \taint{}-based scheme did not lead to a high rate of executable leaks, which we suspect to be due to the sheer number of leaks seeded with it.
The high number of leaks (and more importantly, non-executable leaks) seeded by the \taint{}-based scheme is primarily due to its design, \ie, it places one source per method in a class, thus creating a total of \(n\) sources distributed in \(n\) methods. Then, it places \(n\) sinks per source in each method as scope allows. As a result, around \(n^2\) number of sinks are created in total for \(n\) sources in each class.

The superior performance of the \reachability{} and \complexreachability{} schemes over the other two is expected, \ie, as both \scope{}-based and \taint{}-based schemes insert leaks with sources and sinks distributed across methods, thereby creating unreachable sinks at a higher rate. In contrast, the \reachability{} and \complexreachability{} schemes place leaks with the sources and sinks placed together, resulting in a lower number of unreachable sinks, and hence, a higher rate of creating executable mutants.
\section{In-depth Evaluation of Data Leak Detection Tools with \tool}
\label{sec:indepth_eval}

% \add{
% \AMIT{Changes here}
To demonstrate the utility of \tool, we evaluated three prominent data leak detectors with it: FlowDroid~\cite{arf+14}, HornDroid~\cite{cgm16}, and Argus~\cite{wror14}.
% \add{
    Among these, FlowDroid was analyzed in-depth as part of the previous study~\cite{richie18}, whereas we analyze Argus and HornDroid as part of our current, extended study.
% }
We selected FlowDroid and Argus for the in-depth analysis as they are regularly maintained, with multiple publicly available versions, and form a representative sample of the current state-of-the-art.
Additionally, we selected HornDroid as it is the first Android static analysis tool with a formal proof of soundness, making it an interesting case for a sound{\em i}ness evaluation.
Compatibility with our mutant apps and general analysis feasibility were also major factors that influenced tool selection for the in-depth analysis.
Specifically, we avoided tools that had not been updated recently (\ie, were built for outdated Android versions), and hence were incompatible with many of our mutant apps.

\myparagraph{Methodology} Our methodology for evaluating a security tool with \tool is as follows:
First, we analyze executable mutants with the tool being evaluated.
Then, we systematically examine the surviving (\ie, undetected) mutants using the methodology described in Section~\ref{sec:feasibility}, and resolve the undetected mutants to design/implementation flaws.
Using this approach, our analysis of FlowDroid, HornDroid, and Argus led to the discovery of \(25\) unique flaws,
% \add{
among which $13$ were reported in our \usenix paper~\cite{richie18}.
% }
{\em We confirmed that these flaws were undocumented, \ie{} mentioned neither in the tools' corresponding papers or documentation.}
% }

\subsection{Evaluating FlowDroid}
\label{sec:flowdroid_vuln}

\begin{table*}[t]
\centering
\scriptsize
\caption{{\small Descriptions of flaws uncovered$^*$ in FlowDroid
v$2.0$}}\label{tab:flowdroid-vuln-descs}
\vspace{-0.5em}

\def\arraystretch{1.2}
\begin{tabular}{p{3.9cm} p{9.3cm}}
\toprule
\multicolumn{0}{l}{\textbf{ID: Flaw Name}}   & \multicolumn{1}{l}{\textbf{Description}} \\
% \midrule
\toprule
{} & {\textbf{FC1: Missing Callbacks}}\\
\bottomrule

F1: DialogFragmentShow          &  FlowDroid misses the {\sf DialogFragment.onCreateDialog()} callback registered by {\sf DialogFragment.show()}.\\
% \hline
\midrule

F2: PhoneStateListener          & FlowDroid does not recognize the {\sf onDataConnectionStateChanged()} callback for classes extending the {\sf PhoneStateListener} abstract class from the telephony package.\\
% \hline
\midrule

F3: NavigationView            & FlowDroid does not recognize the {\sf onNavigationItemSelected()} callback of classes implementing the interface {\sf NavigationView.OnNavigationItemSelectedListener}.  \\
% \hline
\midrule

F4: SQLiteOpenHelper           & FlowDroid misses the {\sf onCreate()} callback of classes extending {\sf android.database.sqlite.SQLiteOpenHelper}. \\
% \hline
\midrule

F5: Fragments                 & FlowDroid 2.0 does not model Android
Fragments correctly. We added a patch, which was promptly accepted.
However, FlowDroid 2.5 and 2.5.1 remain affected. We investigate this further in the next section. \\

\toprule
% \multicolumn{2}{l}{\textbf{FC2: Missing Implicit Calls}} \\
{} & {\textbf{FC2: Missing Implicit Calls}}\\

% \Xhline{2\arrayrulewidth}
% \midrule
\midrule

F6: RunOnUIThread               & FlowDroid misses the path to {\sf Runnable.run()} for Runnables passed into {\sf Activity.runOnUIThread()}.\\
% \hline
\midrule

F7: ExecutorService             & FlowDroid misses the path to {\sf Runnable.run()} for Runnables passed into {\sf ExecutorService.submit()}. \\
% \Xhline{2\arrayrulewidth}
\toprule

% \multicolumn{2}{l}{\textbf{FC3: Incorrect Modeling of Anonymous Classes}} \\
{} & {\textbf{FC3: Incorrect Modeling of Anonymous Classes}}\\

% \Xhline{2\arrayrulewidth}
\midrule

F8: ButtonOnClickToDialogOnClick & FlowDroid does not recognize the {\sf onClick()} callback of {\sf DialogInterface.OnClickListener} when instantiated within a Button's {\sf onClick=``method\_name''} callback defined in XML. FlowDroid will recognize this callback if the class is instantiated elsewhere, such as within an Activity's {\sf onCreate()} method.  \\
% \hline
\midrule

F9:  BroadcastReceiver           & FlowDroid misses the {\sf onReceive()} callback of a BroadcastReceiver implemented programmatically and registered within another programmatically defined and registered BroadcastReceiver's {\sf onReceive()} callback.    \\
% \Xhline{2\arrayrulewidth}
% \bottomrule
\toprule
% \multicolumn{2}{l}{\textbf{FC4: Incorrect Modeling of Asynchronous Methods}} \\
{} & {\textbf{FC4: Incorrect Modeling of Asynchronous Methods}}\\

% \Xhline{2\arrayrulewidth}
\midrule

F10: LocationListenerTaint       & FlowDroid misses the flow from a source in the {\sf onStatusChanged()} callback to a sink in the {\sf onLocationChanged()} callback of the {\sf LocationListener} interface, despite recognizing leaks wholly contained in either.\\
% \hline
\midrule
F11: NSDManager                  & FlowDroid misses the flow from sources in any callback of a {\sf NsdManager.DiscoveryListener} to a sink in any callback of a {\sf NsdManager.ResolveListener}, when the latter is created within one of the former's callbacks. \\
% \hline
\midrule
F12: ListViewCallbackSequential &  FlowDroid misses the flow from a source to a sink within different methods of a class obtained via {\sf AdapterView.getItemAtPosition()} within the {\sf onItemClick()} callback of an {\sf AdapterView.OnItemClickListener}.  \\
% \hline
\midrule
F13: ThreadTaint                & FlowDroid misses the flow to a sink within a {\sf Runnable.run()} method started by a Thread, only when that Thread is saved to a variable before {\sf Thread.start()} is called. \\
% \Xhline{2\arrayrulewidth}
\bottomrule
\end{tabular}
\begin{flushleft}
    {\footnotesize * reported in our \usenix paper~\cite{richie18} }
\end{flushleft}
\vspace{-2em}
\end{table*}

%%%%%%%%%%%%%%%%%%%%
%%%%%%%%%%%%%%%%%%%%%%%

% \AMIT{Changes here}
FlowDroid was introduced by Artz et al.~\cite{arf+14} in 2014 as a data leak detection tool for Android.
It models the Android life-cycle to handle callbacks invoked by the Android Framework to perform information flow analysis and data leak detection.
Furthermore, it applies context, flow, field, and object-sensitivity to reduce the number of false positive sensitive data leaks the tool detects. FlowDroid has been cited over 1,400 times, and the tool has been continuously maintained since 2014, which motivated its analysis.

For our in-depth analysis, we evaluated FlowDroid v2.0, which was the version available
% \add{
during our \usenix study~\cite{richie18}, using the \countinsertoldleak mutants originally created from our first seven base apps (\ie, apps 01-07), leveraging the \reachability{}, \complexreachability{} and \taint{} schemes.
% }
In this extension, we used the substantially improved version of \tool to create \countinsertnewleak{} mutants (\countexecutablenewleak executable) from the remaining eight apps to evaluate HornDroid and Argus, using all four mutation schemes (\ie, including the additional \scope{} scheme developed in this extension).
This separation is mainly due to the order in which the tools were evaluated, and the longitudinal nature of our study.
Moreover, this two-phase evaluation with disjoint sets of mutants led to the discovery of flaws in all three tools, which indicates that \tool may be effective at revealing flaws in security tools, irrespective of the apps used as input.
% }

\myparagraph{Results}
Out of the \countinsertoldexecutableleak{} mutants that we analyzed using FlowDroid, \flowdroidnotdetected{} were undetected.
On analyzing the undetected mutants, we discovered $13$ unique flaws in FlowDroid, demonstrating that \tool can be effectively used to find problems that can be resolved to flaws (\ref{rq:findings}).
Using the approach from Section~\ref{sec:feasibility}, we needed less than one hour to isolate a flaw from the set of undetected mutants,  in the worst case.
In the best case, flaws were found in a matter of minutes, demonstrating that the amount of manual effort required to quickly find flaws using \tool{} is minimal (\ref{rq:feasible}).
We provide descriptions of the flaws discovered in FlowDroid in \autoref{tab:flowdroid-vuln-descs}.

We have reported these flaws and are working with the FlowDroid developers to resolve them.
In fact, we developed two patches~\cite{fdroid_git} to correctly implement Fragment support (\ie{} F5 in Table~\ref{tab:flowdroid-vuln-descs}), which were accepted by developers.
To gain insight about the practical challenges faced by static analysis tools, and their design flaws, we further categorize the flaws into the following classes:

\myparagraph{FC1: Missing Callbacks} The security tool (\eg, FlowDroid) did not  recognize some callback method(s) and will not find leaks placed within them. Tools that use lists of APIs or callbacks are susceptible to this problem, as prior work has demonstrated that the generated list of callbacks {\sf (1)} may not be complete, and {\sf (2)} or may not be updated as the Android platform evolves. We observed both such cases in our analysis of FlowDroid. That is, {\sf DialogFragments} was added in API 11 {\em before FlowDroid was released}, and {\sf NavigationView} was added after.  These limitations are well-known in the community of researchers at the intersection of program analysis and Android security, and have been documented by prior work~\cite{cfb+15}.  However, \tool helps evaluate the robustness of existing security tools against these flaws and helps in uncovering these undocumented flaws for the wider security audience. Additionally, {\em some of these flaws may not be resolved even after adding the callback to the list}; \eg, {\sf PhoneStateListener} and {\sf SQLiteOpenHelper}, both added in API 1, are not interfaces, but abstract classes. Therefore, adding them to FlowDroid's list of callbacks (\ie, {\sf AndroidCallbacks.txt}) does not resolve the issue.

\myparagraph{FC2: Missing Implicit Call} The security tool did not identify leaks within some method that is implicitly called by another method.  For instance, FlowDroid does not recognize the path to {\sf Runnable.run()} when a Runnable is passed into the {\sf ExecutorService.submit(Runnable)}. The response from the developers indicated that this class of flaws was due to an unresolved design challenge in Soot's~\cite{vcg+99} SPARK algorithm, upon which FlowDroid depends.  This limitation is also generally well known within the program analysis community~\cite{cfb+15}. However, the documentation of this gap in analysis, thanks to \tool, would certainly benefit researchers in the wider security community.

\myparagraph{FC3: Incorrect Modeling of Anonymous Classes} The security tool did not detect data leaks expressed within an anonymous class. For example, FlowDroid did not recognize leaks in the {\sf onReceive()} callback of a dynamically registered {\sf BroadcastReceiver}, which is implemented within another dynamically registered BroadcastReceiver's {\sf onReceive()} callback. It is important to note that finding such complex flaws is only possible due to \tool's semi-automated mechanism and may be rather prohibitive for an entirely manual analysis.

\myparagraph{FC4: Incorrect Modeling of Asynchronous Methods} The security tool did not recognize a data leak whose source and sink are called within different methods that are asynchronously executed. For instance, FlowDroid did not recognize the flow between data leaks in two callbacks (\ie,\ {\sf onLocationChanged} and {\sf onStatusChanged}) of the {\sf LocationListener} class, which the adversary may cause to execute sequentially (\ie, as our EE confirmed).

Apart from {\bf FC1}, which may be patched with limited effort, the other three  categories of flaws may require a significant amount of research effort to resolve. However, documenting them is critical to increase awareness of real challenges faced by Android static analysis tools.
%% ARGUS
% \add{
\subsection{Evaluating Argus}
\label{sec:argus_vuln}
% \AMIT{changes here}
\begin{table*}[htp]
\centering
\scriptsize
\caption{{\small Descriptions of flaws uncovered in Argus v$3.1.2$.}}
 \label{tab:argus-vuln-descs}
\vspace{-0.5em}
\def\arraystretch{1.2}
\begin{tabular}{p{4cm} p{9cm}}
\toprule

\multicolumn{0}{l }{\textbf{ID: Flaw Name}}   & \multicolumn{1}{l}{\textbf{Description}} \\
\midrule
{} & {\textbf{FC4: Incorrect Modeling of Asynchronous Methods
}
}\\
\midrule

F14: FragmentEventToExternalMethod  & Fragment declared within an Activity class requires its click event listening methods to be defined in the Activity class. When we placed source in such click event listening method and the sink in an Activity method callable by, Argus missed the leak.
\\
\midrule

F15: FragmentCrossClickEventListeners & In similar construction of F14, when source and sink are distributed across click event listener methods connected to Fragment GUI components, Argus missed the leak.\\
\midrule

F16: OnCreateFragmentClickEventListener &
        Android Activity lifecycle method  {\sf onCreate} can be used to create a source for leak. If it is then leaked through a method called by a fragment component click event listening method, it is undetected by Argus.
 \\
\midrule

{} & {\textbf{FC1: Missing Callbacks}}\\
\midrule

F17: FragmentClickEventListener &
            When sink and source are placed in an event listener method defined in an Activity class are coupled with components in a Fragment through relevant XML resource file, Argus misses the leak.
            In similar construction to F14, if both source and sink are placed in event listener methods for fragment components, Argus does not report it.
        \\
\midrule

% AD5
F18: RecyclerViewHolder & A \recyclerviewholder abstract class is used to describe each item within a \recycler. This is required to be extended when used inside the extending class of \recycleradapter. When leak source and sink are placed within the scope of the class implementing the \recyclerviewholder, Argus is unable to detect it.\\
\midrule

F19: RecyclerViewConstructor & This flaw is similar to F18, but where the leak source and sink are placed within the constructor of the class extending \recyclerviewholder. Argus is unable to detect the placed leak in this scenario.\\
\midrule

F20: RecyclerOnCreateViewHolder & The class extending \recycleradapter implements the abstract method \textsf{OnCreateViewHolder\xspace}, when \textsf{ViewHolder\xspace} needs to represent an item. If a leak is placed within \textsf{OnCreateViewHolder\xspace}, Argus's analysis can't detect it.
\\
\midrule

F21: RecyclerViewOnBindViewHolder & Similar to RecyclerCreateViewHolder, the class extending \recycleradapter implements the abstract method \textsf{onBindViewHolder\xspace}, when \textsf{ViewHolder\xspace} needs to display the data at the specified position. If a leak is placed within \textsf{onBindViewHolder\xspace}, Argus's analysis cannot detect it.
\\
\midrule
% AD9
F22: RecyclerViewGetItemCount & \textsf{getItemCount\xspace} is an abstract method required to be overridden to return the total number of items bound in \recycler. This method is placed within a class extending \recycleradapter. Argus is unable to find a leak if the source and sink are placed within \textsf{getItemCount\xspace}.
\\
\bottomrule
\end{tabular}
\vspace{-1.5em}
\end{table*}

% \add{
As part of this extended study, we
% }
% We
evaluated Argus v\(3.1.2\) (\ie, the latest version at the time of our analysis) with \countexecutablenewleak executable mutants generated from the 8 new base apps, \ie, apps 08-15.
Further, we analyzed the \argusnotdetected{} uncaught mutants (\ie, the leaks not detected by Argus) using the methodology previously described in Section~\ref{sec:feasibility}.
% \add{
Through the addition of 8 new base apps, we aim to add diversity to the created mutants, as these new apps are likely to encompass a new set of development practices that might impact our findings.
Therefore, this new set of apps might also help us find new flaws through \tool due to the additional, diverse mutations created. While we do not consider the original apps from our USENIX'18 study for Argus and HornDroid, we consider this to be a balanced trade-off as we later study the propagation of flaws from both old and new sets of apps across tools.
% }

\myparagraph{Results} Through analyzing the uncaught mutants, we discovered \(9\) unique flaws in Argus, as shown and classified in Table~\ref{tab:argus-vuln-descs}.
Note that these flaws are \textit{separate from} the ones found in FlowDroid (Section~\ref{sec:flowdroid_vuln}).
% \add{
This demonstrates that the flaws found from our \usenix study~\cite{richie18} were not inherently coupled to the base apps chosen, and that \tool can be effective with different sets of apps representing a reasonable level of diversity.
    % }
A description of each flaw is provided in Table~\ref{tab:argus-vuln-descs}.
Our project repository provides minimal APKs representing the flaws in Argus~\cite{appendix}.

Of the \(9\) flaws, \(3\) are based on Fragment usage and are of the FC4 flaw class, while the remaining \(6\) fall in FC1.
Note that while we found fragment-based flaws in FlowDroid, the 4 fragment related flaws discovered for Argus are independent, although they could be interpreted as variants of those found in FlowDroid.
The flaws discovered in Argus are described as follows:

\myparagraph{Argus flaws in FC4}
Similar to FlowDroid, Argus struggles with identifying leaks in fragments.
We observe that most of these problems are at the design-level, and occur because Argus does not track flows between the GUI components defined in fragments.
To elaborate, Argus misses asynchronous flows between GUI components defined in the relevant XML files for fragments and the corresponding click-event listeners.
All of our fragment-based flaws (F14-16) exploit this design-gap in Argus to avoid detection.

\myparagraph{Argus flaws in FC1}
Of the remaining \(6\) flaws we discovered in Argus, \(5\) are based on \recycler{} widgets.
The \recycler{} is used to display a collection of data within a limited, scrolling, window.
Each \recycler{} widget implementation includes classes extending \recyclerviewholder{} and \recycleradapter{}.
Furthermore, these classes contain several abstract methods which must be implemented, namely {\sf onCreateViewHolder}, {\sf onBindViewHolder} and {\sf getItemCount}.
Our analysis reveals that Argus fails to detect leaks placed in any of these components and methods.
In addition, Argus does not detect flaws when leaks are placed in click event listeners statically connected to fragment components via XML resource files.
This demonstrates that further work is required to analyze relations in between methods not only through source code, but resource files as well.
Summary of these flaws are tabulated in Table~\ref{tab:argus-vuln-descs}, and the example minimal APKs, each exhibiting a single flaw are available in our online appendix~\cite{appendix}.

\subsection{Evaluating HornDroid}
\label{sec:horndroid_vuln}
HornDroid was proposed by Calzavara et al.~\cite{cgm16} as the first static analysis tool for Android with a formal proof of soundness.
This unique attribute motivated our choice of HornDroid for in-depth evaluation
% \add{
    in this extended study.
% }.
HornDroid abstracts Android applications as a set of Horn clauses to formulate security properties, which can then be processed by Satisfiability Modulo Theories (SMT) solvers.

HornDroid's formal proof and over-approximation require far more resources than the other tools we evaluate, \ie, as stated in the HornDroid paper, the authors tested HornDroid on a server with 64 multi-thread cores and 758 Gb of memory, although they reported that the most memory utilization was around 10 GB.
To match this maximum 10 GB memory utilization, we conducted our study on a server with 32GB of RAM and 8 cores.
When analyzing a mutated app with HornDroid, we set a time-out of 36 hours, after which we would abort the analysis and report whatever mutants were caught until that time.

\begin{table*}
\centering
\scriptsize
\caption{{\small Descriptions of flaws uncovered in HornDroid.}}
\vspace{-0.5em}
\def\arraystretch{1.2}
\begin{tabular}{p{4cm} p{9cm}}
\toprule
\multicolumn{0}{l}{\textbf{ID: Flaw Name}}   & \multicolumn{1}{l}{\textbf{Description}} \\
\midrule

{} & {\textbf{FC5: Android Lifecycle Callbacks}}\\
\midrule
% Fragment variant 0 origin
F23: OnCreateFragmentConstructor & HornDroid does not detect a leak if the source is placed in the \textsf{onCreate} method of an activity, and the sink is placed in the constructor of the fragment within the activity. 
\\ 
\midrule
% Fragment variant 0
F24: ActivityOncreate & When a leak is placed in the \textsf{onCreate} method of Activity, i.e. both source and sink are placed in the method, HornDroid is unable to detect the leak. 
\\ 
\midrule
% Fragment variant 1
F25: FragmentOnCreateView & When the source of the leak is placed at the \textsf{onCreate} method of activity class, and the sink at the \textsf{onCreateView} method of the fragment class, HornDroid does not detect the leak.
\\
\bottomrule

\end{tabular}
\label{tab:horn-vuln-descs}
\vspace{-1.5em}
\end{table*}

% }
\myparagraph{Results} We were not able to analyze many of our mutated apps using HornDroid.
Specifically, out of the \countmutantappsfromlatereight{} mutated APKs (\ie, created after mutating the base APKs 08-15), HornDroid could only analyze 4 without crashing or timing out.
This was in spite of us taking care to compile the APKs with the API level that HornDroid was built to analyze, \ie, API 19.
This outcome is unfortunately not surprising; indeed, prior work on analyzing the feasibility of Android static analysis tools has reported similar results for research prototypes~\cite{rbg+16}.
As a result, HornDroid successfully analyzed \horndroidtestedagainst{} executable data leaks, out of which, it caught \horndroiddetected{}, leaving \horndroidnotdetected{} undetected mutants for further analysis.
We discovered 3 flaws from analyzing these undetected mutants, as shown in Table~\ref{tab:horn-vuln-descs}.

All of the 3 flaws (F23-F25) we discovered are related to the lack of appropriate support for fragments; however, unlike prior fragment-related flaws, these flaws are generally centered around the Android lifecycle methods for fragments and activities.
These flaws could broadly fit in FC1; however, they demonstrate a more fundamental, design-level lack of support for fragments, as HornDroid fails to detect leaks in even basic lifecycle methods (\eg, Fragment onCreateView).
Hence, we create a new flaw class to represent such flaws, {\em FC5: Android Lifecycle Callbacks}, as a special variant of FC1, described as follows:

\myparagraph{FC5: Android Lifecycle Callbacks} This class incorporates flaws in detecting leaks in fundamental lifecycle methods, such as onCreate.
To elaborate, there are only six lifecycle call back methods in total for any Android activity~\cite{android-lifecycle}, where {\sf onCreate} is the only one considered mandatory.
It is also the first method to be called when the Activity is initialized \ie{} it is the starting point of the Activity.
Hence, analyzing such callbacks is imperative for a practical analysis of data flows within the app.
However, HornDroid fails to detect leaks in three specific instances of lifecycle callbacks for activities and fragments (F23-F25), which is concerning, as the tool strongly claims soundness, \ie, quoted as follows: \textit{"In order to support a sound analysis of fragments, HornDroid over-approximates their life-cycle by executing all the fragments along with the containing activity in a flow-insensitive way".}
% }
% change ends here

% \add{
\subsection{Effectiveness of Individual Schemes in Finding Flaws}
\label{sec:flaws_from_scheme}
% \AMIT{changes here}
\begin{table*}[t]
    \centering
    \scriptsize
    \caption{{\small Impact of Operator Placement Approaches in Finding Flaws.$^*$}}
    \label{tab:flaw-back-scheme}
	\vspace{-0.5em}
\begin{tabular*}{\textwidth}{l @{\extracolsep{\fill}}  c c c c }
\toprule

{\bf Flaw ID}	& {\bf Reachability} & {\bf Complex-Reachability} & {\bf
    Taint} & {\bf Scope}  \\
\midrule
\multicolumn{5}{c}{Flaws found from FlowDroid v{$2.0$}}\\
\midrule
F1	 & \checkmark{}	& \checkmark{}	& \checkmark{}	& \checkmark{} \\
F2	 & \checkmark{}	& \checkmark{}	& \checkmark{}	& \checkmark{} \\
F3	 & \checkmark{}	& \checkmark{}	& \checkmark{}	& \checkmark{} \\
F4	 & \checkmark{}	& \checkmark{}	& \checkmark{}	& \checkmark{} \\
F5	 & \checkmark{}	& \checkmark{}	& \checkmark{}	& \checkmark{} \\
F6	 & \checkmark{}	& \checkmark{}	& \checkmark{}	& \checkmark{} \\
F7	 & \checkmark{}	& \checkmark{}	& \checkmark{}	& \checkmark{} \\
F8	 & \checkmark{}	& \checkmark{}	& \checkmark{}	& \checkmark{} \\
F9	 & \checkmark{}	& \checkmark{}	& \checkmark{}	& \checkmark{} \\
F10 & -	        & -	            & \checkmark{}	& \checkmark{} \\
F11 & -	        & -	            & \checkmark{}	& \checkmark{} \\
F12 & -	        & -	            & \checkmark{}	& \checkmark{} \\
F13 & -	        & -	            & \checkmark{}	& \checkmark{} \\
\midrule
\multicolumn{5}{c}{Flaws found from Argus v{$3.1.2$}}\\
\midrule
F14	 & -	        & -	            & \checkmark{}	& \checkmark{} \\
F15	 & -	        & -	            & \checkmark{}	& \checkmark{} \\
F16	 & -	        & -	            & \checkmark{}	& \checkmark{} \\
F17	 & \checkmark{}	& \checkmark{}	& \checkmark{}	& \checkmark{} \\
F18	 & \checkmark{}	& -	            & -	            & -            \\
F19	 & \checkmark{}	& \checkmark{}	& \checkmark{}	& \checkmark{} \\
F20	 & \checkmark{}	& \checkmark{}	& \checkmark{}	& \checkmark{} \\
F21	 & \checkmark{}	& \checkmark{}	& \checkmark{}	& \checkmark{} \\
F22	 & \checkmark{}	& \checkmark{}	& \checkmark{}	& \checkmark{} \\
\midrule
\multicolumn{5}{c}{Flaws found from HornDroid}\\
\midrule
F23	 & \checkmark{}	& \checkmark{}	& \checkmark{}	& \checkmark{} \\
F24	 & \checkmark{}	& \checkmark{}	& \checkmark{}	& \checkmark{} \\
F25	 & -	        & -	            & -	            & \checkmark{} \\

\bottomrule

\end{tabular*}
\begin{flushleft}
    {\footnotesize $^*$A ``\checkmark''
    indicates flaw may be resolved through relevant operator placement approach, whereas ``{\sf -}'' indicates it may not. }
\end{flushleft}
\vspace{-1.5em}
\end{table*}

As described in Section~\ref{sec:mutation-scheme}, \tool uses four different mutation schemes to place operators in an app.
These schemes may sometimes create overlapping mutants wherein operators are placed in the same position according to two or more schemes.
% \add{
    As part of this extended study,
% }
we trace the flaws we found back to the operator placement strategies described in Section~\ref{sec:reachability-scheme} and~\ref{sec:goal-scheme} to determine the usefulness of these approaches in finding flaws (\ref{rq:effectiveness}).

As shown in \autoref{tab:flaw-back-scheme}, of the 25  flaws, 24 could be discovered using the scope-based scheme, and 22 using the taint-based scheme.
In addition, we found one flaw (\ie, F18) that could only be reached using the reachability-based scheme.
Another interesting observation is that both the \reachability{} and \complexreachability{} schemes are similar in terms of usefulness in finding flaws, which may be a positive indicator of a general reachability-based approach over more complex strategies.

Based on the data in \autoref{tab:flaw-back-scheme} alone, it may seem that only using the scope-based scheme is a viable option.
However, recall from Section~\ref{sec:executing_tools} that mutation schemes may also display widely different performance in terms of creating {\em executable} mutants.
As a result, using a single scheme may improve flaw detection, but may also generate a tremendous overhead in terms of non-executing mutants.
Furthermore, even if the same mutation/leak placement can be achieved using multiple schemes, the common placement may help the researcher or tool developer resolve the flaw behind an uncaught mutant faster, \ie, by helping them see the common or different factors (\ie, among the schemes).
For example, the \complexreachability{} scheme makes the path between source and sink indirect while residing within the same scope, whereas the {\sf reachability} scheme places the source and sink at the same location without establishing any indirect path.
Failure to detect the leak placed through \complexreachability{} would indicate that the flaw is due to the indirect path, rather than the location of the leak.
% }
% change ends here

% propagation study
% \add{
\section{Flaw Propagation Study}
\label{sec:propagation_study_pre}
The objective of this study is to determine whether flaws from one tool propagate  to other tools, which are either implemented directly on top of the original tool, or inherit certain design attributes (\ref{rq:parent_inherit}).
To carry out this study, we utilized the minimal APKs created for each of our 25 flaws, and analyzed them with other tools built with the same security goal (\ie, other data leak detectors excluding the tools from which the specific flaws were discovered).
Each minimal APK contains only one type of flaw, \ie, we built 25 APKs, one for each flaw mentioned in Tables \ref{tab:flowdroid-vuln-descs}, \ref{tab:argus-vuln-descs} and \ref{tab:horn-vuln-descs}.

Moreover, in order to prevent tools from crashing due to backwards compatibility issues, we built several versions of the minimal APKs from the same code base, varying the SDK versions, as well as the build tools (\ie, Android Studio and Gradle, vs building manually using Android SDK tools).
This was done because APK building procedure has changed over the years, and as a result, many of the studied tools, which are well over 4-5 years old, would break for apps built using the latest build configuration and procedure.
For example, Android Studio's Gradle based building overrides the target SDK version defined in the AndroidManifest.xml file, and also uses newer versions of build tools and target platforms.
We discovered that decompilers such as dare~\cite{Octeau:FSE2012} used in some of the tools we analyzed do not function correctly when analyzing such builds (\ie, either crash, or run into infinite loops).
Considering these factors, we try our best to customize minimal apks for individual tools, in order to minimize crashes or timeouts, and only report results from variants of minimal APKs that worked., \ie, resulted in a successfully completed analysis.

\subsection{Propagation of FlowDroid's Flaws (F1-F13)}
\label{sec:vuln_propagation}
\begin{table*}[t]
\scriptsize
\centering
\caption{{\small Analysis of the propagation of flaws in FlowDroid 2.0 to other data leak detectors.$^*$}}
\vspace{-0.5em}
\label{tbl:vuln_table}
\setlength{\tabcolsep}{3pt}

\begin{tabular*}{\textwidth}{l @{\extracolsep{\fill}} c c c c c c c c c c}
\toprule

\bf {Flaw ID}\hspace{0.1cm} &
{\bf FD v2.0}	&
{\bf FD v2.5} &
{\bf FD v2.5.1} &
{\bf FD v2.7.1} &
{\bf Argus} &
{\bf BlueSeal} &
{\bf DidFail} &
{\bf DroidSafe} &
{\bf HornDroid} &
{\bf IccTA}
\\
\midrule

F1		& \checkmark  & \checkmark & \checkmark & \checkmark & {\sf x}    & {\sf x}    & \checkmark & {\sf x}    & \checkmark &   \checkmark\\
F2		& \checkmark  & \checkmark & \checkmark & \checkmark & {\sf x}    & {\sf x}    & \checkmark & {\sf x}    & \checkmark &   \checkmark\\
F3		& \checkmark  & \checkmark & \checkmark & \checkmark & \checkmark & -          & \checkmark & -          & -          &   \checkmark\\
F4		& \checkmark  & \checkmark & \checkmark & \checkmark & \checkmark & {\sf x}    & \checkmark & {\sf x}    & \checkmark &   \checkmark\\
F5		& \checkmark  & \checkmark & \checkmark & \checkmark & \checkmark & \checkmark & \checkmark & -          & \checkmark &   \checkmark\\
F6		& \checkmark  & \checkmark & \checkmark & \checkmark & \checkmark & {\sf x}    & \checkmark & {\sf x}    & \checkmark &   \checkmark\\
F7		& \checkmark  & \checkmark & \checkmark & \checkmark & \checkmark & {\sf x}    & \checkmark & {\sf x}    & \checkmark &   \checkmark\\
F8 	    & \checkmark  & \checkmark & \checkmark & \checkmark & {\sf x}    & {\sf x}    & \checkmark & \checkmark & {\sf x}    &   \checkmark\\
F9		& \checkmark  & \checkmark & \checkmark & \checkmark & {\sf x}    & {\sf x}    & \checkmark & {\sf x}    & {\sf x}    &   \checkmark\\
F10	    & \checkmark  & \checkmark & \checkmark & \checkmark & {\sf x}    & {\sf x}    & \checkmark & {\sf x}    & {\sf x}    &   \checkmark\\
F11	    & \checkmark  & \checkmark & \checkmark & \checkmark & \checkmark & {\sf x}    & \checkmark & {\sf x}    & {\sf x}    &   \checkmark\\
F12     & \checkmark  & \checkmark & \checkmark & \checkmark & {\sf x}    & {\sf x}    & \checkmark & {\sf x}    & {\sf x}    &   \checkmark\\
F13	    & \checkmark  & \checkmark & \checkmark & \checkmark & {\sf x}    & {\sf x}    & \checkmark & {\sf x}    & {\sf x}    &   \checkmark\\

\bottomrule
\end{tabular*}
\begin{flushleft}
    {\footnotesize $^*$ Note that
a ``$-$'' indicates tool crash with the minimal APK, a ``\checkmark''
indicates presence of the flaw, and a ``{\sf x}'' indicates absence; {\sf FD}* = {\sf FlowDroid.}}
\end{flushleft}
\vspace{-1.5em}
\end{table*}

To determine if the flaws present in FlowDroid are also present in other data leak detectors, as well as tools that inherit it, we checked if the newer release versions of FlowDroid (\ie, v$2.5$, v$2.5.1$, v$2.7.1$), as well as $6$ other tools (\ie, Argus, DroidSafe, IccTA, BlueSeal, HornDroid, and DidFail), are susceptible to any of the flaws discussed in Table~\ref{tab:flowdroid-vuln-descs}.
% \add{
Among these, flaw propagation across versions of FlowDroid was done for v$2.5$, v$2.5.1$ in our original \tool{} study. As part of this extended study, due to availability of FlowDroid v$2.7.1$, we include it.
% }

\myparagraph{Results} Table~\ref{tbl:vuln_table} provides an overview of the propagation of F1-F13. In the Table~\ref{tbl:vuln_table}, all the versions of FlowDroid are susceptible to the flaws discovered from our analysis of FlowDroid v2.0. Note that while we fixed the {\sf Fragment} flaw and our patch was accepted to FlowDroid's codebase, the latest releases of FlowDroid (\ie, v$2.5$, v$2.5.1$, and v$2.7.1$) still seem to have this flaw. We have reported this issue to the developers.

A significant observation from Table~\ref{tbl:vuln_table} is that the tools  that directly inherit FlowDroid (\ie, IccTA, DidFail) are similarly flawed as FlowDroid. This is especially true when the tools do not augment FlowDroid in any manner, and use it as a black box (\ref{rq:parent_inherit}). On the contrary, Argus, which is motivated by FlowDroid's design, but augments it on its own, does not inherit as many of FlowDroid's flaws.

BlueSeal, HornDroid, and DroidSafe use a significantly different methodology, and are also not susceptible to many of \tool's uncovered flaws for FlowDroid. Interestingly, BlueSeal and DroidSafe are similar to FlowDroid in that they use Soot to construct a control flow graph and rely on it to identify paths between sources and sinks. However, BlueSeal and DroidSafe both augment the graph in novel ways, and thus do not exhibit most of the flaws found in FlowDroid.

Finally, it is important to note that our analysis does not imply that FlowDroid is weaker than the tools\add{,} which have fewer flaws in Table~\ref{tbl:vuln_table}. However, it does indicate that the flaws discovered may be typical of the design choices made in FlowDroid and inherited by the tools such as IccTA and DidFail.

% \add{

\begin{table*}
\scriptsize
\centering
\caption{{\small Analysis of the propagation of flaws in Argus 3.1.2 to other data leak detectors.$^*$}}
\label{tbl:vuln_table_argus}
\vspace{-0.5em}
\setlength{\tabcolsep}{3pt}
\begin{tabular*}{\textwidth}{l @{\extracolsep{\fill}}  c c c c c c c c c c c}
\toprule
{\bf Flaw ID}\hspace{0.2cm} &

    {\bf Argus} &
    \multicolumn{1}{c }{{\bf FD*~v2.5.1}} &
    \multicolumn{1}{c }{~{\bf FD* v2.6}} &
    \multicolumn{1}{c }{{\bf FD* v2.6.1}} &
    \multicolumn{1}{c}{{\bf FD* v2.7}} &
    \multicolumn{1}{c }{{\bf FD* v2.7.1}} &
    % \multicolumn{1}{c }{{\bf HornDroid}} &
    \multicolumn{1}{c}{{\bf BlueSeal}} &
    \multicolumn{1}{c }{{\bf DidFail}} &
    \multicolumn{1}{c}{{\bf DroidSafe}} &
    % \multicolumn{1}{c}{{\bf BlueSeal}} &
    \multicolumn{1}{c }{{\bf HornDroid}} &
    \multicolumn{1}{c}{{\bf IccTA}}\\

\midrule

F14 & \checkmark & \checkmark  & -          & \checkmark &  -          &  \checkmark & \checkmark  & -       & \checkmark  & \checkmark & \checkmark\\

F15 & \checkmark & \checkmark  & -          & \checkmark &  -          &  \checkmark & \checkmark  & -       & \checkmark  & \checkmark & \checkmark\\
F16 & \checkmark & \checkmark  & -          & \checkmark &  -          &  \checkmark & \checkmark  & -       & \checkmark  & \checkmark & \checkmark\\
F17 & \checkmark & \checkmark  & -          & \checkmark &  -          &  \checkmark & \checkmark  & -       & \checkmark  & \checkmark & \checkmark\\
F18 & \checkmark & \checkmark  & \checkmark & \checkmark &  \checkmark &  \checkmark &  -          & -       & -           & \checkmark &  \checkmark\\
F19 & \checkmark & \checkmark  & \checkmark & \checkmark &  \checkmark &  \checkmark &  -          & -       & -           & \checkmark &  \checkmark\\
F20 & \checkmark & \checkmark  & \checkmark & \checkmark &  \checkmark &  \checkmark &  -          & -       & -           & \checkmark & \checkmark\\
F21 & \checkmark & \checkmark  & \checkmark & \checkmark &  \checkmark &  \checkmark &  -          & -       & -           & \checkmark &  \checkmark\\
F22 & \checkmark & \checkmark  & \checkmark & \checkmark &  \checkmark &  \checkmark &  -          & -       & -           & \checkmark & \checkmark\\

\bottomrule
\end{tabular*}
\begin{flushleft}
    {\footnotesize $^*$ Note that
a ``$-$'' indicates tool crash with the minimal APK, a ``\checkmark''
indicates presence of the flaw, and a ``{\sf x}'' indicates absence; {\sf FD}* = {\sf FlowDroid.}}
\end{flushleft}
\vspace{-1.5em}
\end{table*}

\subsection{Propagation of Argus's Flaws (F14-F22)}
\label{sec:vuln_propagation_argus}
% \AMIT{Changes here}
% \add{
    Unique to this extended study (\ie, previously not reported in our original \tool paper~\cite{richie18}), we
% }
% We
analyzed the minimal APKs developed for F14-F22 with FlowDroid v\(2.5.1\), v\(2.6\), v\(2.6.1\), v\(2.7 \), and v\(2.7.1\); BlueSeal, DroidSafe, DidFail, HornDroid, and IccTA to examine their prevalence.
Note that FlowDroid 2.0 had become unavailable at this point in the study (\ie, deprecated in favor of later versions), which is why we focus on FlowDroid versions from 2.5 to 2.7.1.

\myparagraph{Results} As shown in the Table~\ref{tbl:vuln_table_argus}, flaws found in Argus largely affect FlowDroid, HornDroid and IccTA.
Specifically, recall that while some of the fragment and \recycler-based flaws were missing callbacks that could potentially be fixed with patches, some were design-level gaps in the data flow tracking performed by these tools.
For example, in F14, an event listener reads sensitive data (\ie, the source), and calls another method, which leaks the data (\ie, the sink).
While such event listeners may not be directly called from any of the lifecycle methods in the app, they are likely to be invoked when  user interacts with the GUI components of the app.
The fact that none of the \recycler-based (\ie, F18-F22) or fragment-based (\ie, F14-F17) leaks were detected by any of the tools we evaluated demonstrates the fragility of these tools in face of commonly-used Android GUI abstractions.

DroidSafe and BlueSeal crashed when analyzing the minimal APKs for F18-F22, while DidFail crashed for all flaws, \ie, F14-F22.
These crashes are expected, and primarily occur due to lack of support for newer API.
To elaborate, DroidSafe was specifically built to focus on Android 4.4.1 (API 19), and hence crashes on encountering  \recycler{}, \ie, the set of APIs at the root of F14-F22, which was introduced in Android 5.0 (API 21).
Similarly, BlueSeal customizes an older version of Soot for analysis, which fails when analyzing \recycler{}-based APKs.
Finally, DidFail was built using customized but outdated versions of several other tools, namely FlowDroid, Epicc~\cite{octeauEffectiveIntercomponentCommunication2013}, and dare~\cite{Octeau:FSE2012}, and hence crashes on all the newer flaws.
Note that while tools such as FlowDroid and dare may be separately updated and maintained, the current implementation of DidFail customizes these dependencies to a sufficient degree, which prevents us from simply replacing the dependency with a newer version.
However, we can still infer that since DidFail uses FlowDroid as a component, it is likely to inherit all of FlowDroid's flaws.
\subsection{Propagation of HornDroid's Flaws (F23-F25)}
\label{sec:vuln_propagation_horndroid}
\begin{table*}[t]
\scriptsize
\centering
\caption{{\small Analysis of the propagation of flaws in HornDroid to other data leak detectors.$^*$ }}
\label{tbl:vuln_table_horndroid}
\vspace{-0.5em}
\setlength{\tabcolsep}{3pt}
\begin{tabular*}{\textwidth}{l @{\extracolsep{\fill}} c c c c c c c c c c c}

\toprule
\textbf{Flaw ID}\hspace{0.2cm} &

\textbf{Argus} &
\multicolumn{1}{c }{\textbf{FD*~v2.5.1}} &
\multicolumn{1}{c }{~\textbf{FD* v2.6}} &
\multicolumn{1}{c }{\textbf{FD* v2.6.1}} &
\multicolumn{1}{c }{\textbf{FD* v2.7}} &
\multicolumn{1}{c }{\textbf{FD* v2.7.1}} &
\multicolumn{1}{c }{\textbf{BlueSeal}} &
\multicolumn{1}{c }{\textbf{DidFail}} &
\multicolumn{1}{c}{\textbf{DroidSafe}} &
\multicolumn{1}{c}{\textbf{HornDroid}} &
\multicolumn{1}{c}{\textbf{IccTA}}\\

\midrule
  -   & Argus      & FD2.5.1     & 2.6        & 2.6.1      & 2.7         & 2.7.1       & BlueSeal   & DidFail     & DroidSafe             & HornDroid  & IccTA\\
F23   & x          & x           & -          & x          &  -          &  \checkmark &  x         & -           & \checkmark            & \checkmark & x \\
F24   & x          & x           & -          & x          &  -          &  \checkmark &  x         & -           & \checkmark            & \checkmark & x \\
F25   & x          & \checkmark  & -          & \checkmark &  -          &  \checkmark &  x         & -           & \checkmark            & \checkmark & \checkmark \\

\bottomrule
\end{tabular*}
\begin{flushleft}
    {\footnotesize $^*$ Note that
a ``$-$'' indicates tool crash with the minimal APK, a ``\checkmark''
indicates presence of the flaw, and a ``{\sf x}'' indicates absence; {\sf FD}* = {\sf FlowDroid.}}
\end{flushleft}
\vspace{-1.5em}
\end{table*}

To understand whether the flaws we found in HornDroid propagate to other tools, we prepared minimal APKs based on the found flaws of HornDroid, and analyzed them using Argus, FlowDroid versions v\(2.5.1\), v\(2.6\), v\(2.6.1\), v\(2.7 \), and v\(2.7.1\);  BlueSeal, DroidSafe, DidFail, and IccTA.
% \add{
None of these flaws, nor their propagation, were reported in our original \tool{} paper~\cite{richie18}.
% }

\myparagraph{Results} As shown in the Table~\ref{tbl:vuln_table_horndroid}, we find that Argus did not exhibit any of the fragment-based flaws identified in HornDroid.
This is surprising, considering that HornDroid was susceptible to every single flaw discovered in Argus, as seen previously in Table~\ref{tbl:vuln_table_argus}.
This finding may indicate that Argus may be relatively more sound than HornDroid in practice, in spite of the latter providing a formal proof of soundness.

Furthermore, although FlowDroid's earlier versions (v\(2.5.1\), v\(2.6.1\)) were able to detect leaks F23 and F24, FlowDroid version v\(2.7.1\) was not.
This is not an isolated case: recall that our patch for another fragment flaw F5~\cite{fdroid_git} fixed it in FlowDroid v2.0, but future versions of FlowDroid (\ie, v2.5 onwards) still exhibit the flaw, as seen in Table~\ref{tbl:vuln_table}.
Interestingly, we note that F23 and F24 are absent in IccTA, an approach that relies on FlowDroid. This is because IccTA uses an older version of FlowDroid as a component, which is resistant to these flaws, and hence, remains unaffected as well.
Moreover, F25, a fragment-based flaw, is exhibited by all versions of FlowDroid.
This finding is an indicator of the lack of systematic fragment support in FlowDroid, as well as other major tools, when in fact fragments are a widely-used GUI element that may contain data leaks.

Finally, the propagation study also allows us to derive certain general conclusions regarding the quality of the tools studied.
First, we conclude that all of the tools we analyzed are incapable of finding leaks that are exhibited in \recycler{}s.
Second, we find evidence to suggest that Argus is relatively better for detecting fragment-based leaks, relative to HornDroid and FlowDroid.
This may seem counterintuitive considering there was only one fragment related flaw (F5) in FlowDroid, two in HornDroid (F23, F25) and four in Argus (F14-17).
However, when we consider the propagation of fragment related flaws as well, Argus is affected by five (F5, F14-17), while HornDroid and FlowDroid (considering its latest release, v2.7.1) are both affected by seven (\ie, both are affected by flaws F5, F14-17, and F23, F25).
Thus, we argue that Argus provides more holistic support for fragments, relative to the other tools studied.

% Effectiveness of mutation schemes
% \input{texfiles/mutation_scheme_effectiveness}

\section{Discussion}
\label{sec:discussion}

\tool has demonstrated efficiency and effectiveness at revealing real  undocumented flaws in prominent Android security analysis tools.  While experts in Android static analysis may be familiar with some of the flaws we discovered (\eg,{} some flaws in FC1 and FC2), we aim to document these flaws for the entire scientific community. Further, \tool{} indeed found some design gaps that were surprising to expert developers; \eg,{} FlowDroid's design does not consider callbacks in anonymous inner classes (flaws 8-9, Table~\ref{tbl:vuln_table}), and in our interaction with the developers of FlowDroid, they acknowledged handling such classes as a non-trivial problem. During our evaluation of \tool{} we were able to glean the following pertinent insights:

% {\color{blue}
\myparagraph{Insight 1} {\em Most mutation schemes are generally effective.}
While certain mutation schemes may be Android-specific, our results demonstrate limited dependence on these configurations.
Out of the 25 flaws discovered using \tool (\ie, both in our \usenix paper~\cite{richie18} as well as this extension), we discover that each mutation scheme is necessary for detecting certain flaws (\ie, which may not be detected with other schemes), as shown in Section~\ref{sec:flaws_from_scheme}.
% }
%}

\myparagraph{Insight 2} {\em Security-focused static analysis tools exhibit  undocumented flaws that require further evaluation and analysis}.  Our results clearly demonstrate that previously unknown security flaws or undocumented design assumptions, which can be detected by \tool{}, pervade existing Android security static analysis tools. Our findings not only motivate the dire need for systematic discovery, fixing and documentation of unsound choices in these tools, but also clearly illustrate the power of mutation based analysis adapted in security context.

\myparagraph{Insight 3} {\em Current tools inherit flaws from legacy tools}. A  key insight from our work is that while inheriting code of the foundational tools (\eg,{} FlowDroid) is a common practice, some of the researchers may not necessarily be aware of the unsound choices they are inheriting as well.  As our study results demonstrate, when a tool inherits another tool directly (\eg, IccTA inherits FlowDroid), all the flaws propagate.

\myparagraph{Insight 4}{\em \space Tools which follow similar design principles but do not have a direct relationship (\eg, inheriting a codebase), have similar flaws.}
Through our experiments and evaluation performed in this extended study,
we effectively demonstrate that FlowDroid, HornDroid, and Argus; three different static analysis tools built independently from each other, but which share similar design principles, can and do exhibit similar flaws.
This indicates that certain unsound decisions or flaws may be tied to the common security goal or could be occurring due to fundamental gaps in the high-level design decisions that are common across the board for such tools.

\myparagraph{Insight 5 }{\em \space Flaws which were not present in previous versions of a static analysis tool can appear in later versions, as the tool evolves.}
As we found in this extended study,
certain flaws found in HornDroid were not present in earlier versions of FlowDroid but appeared in the latest version (Table~\ref{tbl:vuln_table_horndroid}).
This shows that unsound choices can be made at any stage and at any iteration of software life cycle, and further establishes the necessity of automatically and systematically evaluating these tools.
\tool lays the groundwork for the development of such a holistic, dynamic, testing framework.

\myparagraph{Insight 6} {\em As tools, libraries, and the Android  platform evolve, security problems become harder to track down}. Due the nature of software evolution, all the analysis tools, underlying libraries, and the Android platform itself evolve asynchronously. A few changes in the Android API may introduce undocumented flaws in analysis tools. \tool{} handles this fundamental obstacle of continuous change by ensuring that each version of an analysis tool is systematically tested, as we realize while tracking the Fragment flaw in multiple versions of FlowDroid.

\myparagraph{Insight 7} {\em Benchmarks need to evolve with time}.  While manually-curated benchmarks (\eg,{} DroidBench~\cite{arf+14}) are highly useful as a ``first line of defense'' in checking if a tool is able to detect well-known flaws, the downside of relying too heavily on benchmarks is that they only provide a known, finite number of tests, leading to a false sense of security. Due to constant changes (insight \#6) benchmarks are likely to become less relevant unless they are constantly augmented, which requires tremendous effort and coordination. \tool{} significantly reduces this burden on benchmark creators via its suite of extensible and expressive security operators and mutation schemes, which can continuously evaluate new versions of tools. The key insight we derive from our experience building \tool{} is that {\em while benchmarks may check for documented flaws, \tool{}'s true strength is in discovering new flaws.}

\section{Related Work}
\label{sec:relwork}

\tool builds upon the theoretical underpinnings of mutation analysis  from SE, and to our knowledge, is the first work to adapt mutation analysis to evaluate the soundness claimed by security tools. Moreover, \tool adapts mutation analysis to security, and makes fundamental and novel modifications (described previously in Section~\ref{sec:design}).
We now describe prior work in three other related areas:

\myparagraph{Formally Verifying Soundness} While an ideal approach,   formal verification is one of the most difficult problems in computer security.  For instance, prior work on formally verifying apps often requires the monitor to be rewritten in a new language or use verification-specific programming constructs (\eg, verifying reference monitors~\cite{fcds10,vcj+13}, information flows in apps~\cite{mye99,ml00,yys12}), which poses practical concerns for tools based on numerous legacy codebases (\eg, FlowDroid~\cite{arf+14}, CHEX~\cite{llw+12}). Further, verification techniques generally require correctness to be specified, \ie, the policies or invariants that the program is checked against. Concretely defining what is ``correct'' is hard even for high-level program behavior (\eg, making a ``correct'' SSL connection) and may be infeasible for complex static analysis tools (\eg, detecting ``all incorrect SSL connections'').  \tool does not aim to substitute formal verification of static analysis tools; instead, it aims to uncover existing limitations of such tools.
%\AMIT{Changes here}

% {\color{blue}
\myparagraph{Evaluating Static Analyses}
Recently, there has been significant work in the area of experimentally evaluating the features and effectiveness of static analysis techniques.
For instance, Qiu \etal{}~\cite{QWR18} performed a comparative evaluation of precision and runtime performance, among FlowDroid+IccTA, AmanDroid and DroidSafe, by using a common configuration setup, using a benchmark that extends DroidBench~\cite{droidbench} and ICC-Bench~\cite{iccbench}.
Pauck \etal{}~\cite{PBW18} propose the ReproDroid framework that automatically evaluates the effectiveness of Android taint analysis tools using user-labeled ground truth in Android apps.
However, there are fundamental differences in our work, and these related approaches, in terms of the primary goal, scope, and the actual techniques leveraged.

To elaborate, \tool focuses on exhaustively generating security test cases for evaluating Android security tools, and systematically performing in-depth evaluations of such tools to discover gaps in the soundness that directly affect their ability to detect security vulnerabilities such as data leaks.
\tools security focus is evident in our additional efforts towards designing security-focused mutation, and attributing undetected mutants to actual flaws and design choices in tools that affect security.
On the contrary, both prior approaches focus on evaluating either the presence of promised static analysis {\em features}, or the general precision, \ie, with a lack of specific focus on security.
Furthermore, \tool is a holistic, automated, mutation framework that generates thousands of expressive, {\em security-goal-focused test cases} for evaluating security tools, while related work generally relies on handcrafted benchmarks~\cite{QWR18} or  user-specified ground-truth~\cite{PBW18}, which may be sufficient for evaluating features, but not for a thorough security evaluation of tools.
That is, \tool{} \textit{empowers} security researchers to discover flaws without delving into the intricate details of program analysis techniques.
However, we do note that certain aspects of prior work (\eg, the automated bootstrapping of tools in ReproDroid~\cite{PBW18}) are complementary to \tool, and may be incorporated into its pipeline in the future.
Finally, while \tool evaluates soundness, our mutation-based approach may be used to evaluate precision as well, which is a separate research direction that we aim to explore in the future.
The popularity and open-source nature of Android has spurred an  immense amount of research related to examining and improving the security of the underlying OS, SDK, and apps. Recently, Acar \etal have systematized   Android security research \cite{androidsok:sp16}, and we discuss work that introduces static analysis-based countermeasures for Android security issues according to Acar \etals categorization.

Perhaps the most prevalent area of research in Android security  has concerned the permissions system that mediates access to privileged hardware and software resources.  Several approaches have motivated changes to Android's permission model, or have proposed enhancements to it, with goals ranging from detecting or fixing unauthorized information disclosure or leaks in third party applications~\cite{egc+10,arf+14,gcec12,ne13,naej16,xw15,jaf+13} to  detecting over privilege in applications~\cite{fch+11,azhl12,vcc11}. Similarly, prior work has also focused on benign but vulnerable Android applications, and proposed techniques to detect or fix vulnerabilities such as cryptographic API misuse API~\cite{fhm+12,ebfk13,ssg+14,fhp+13} or unprotected application interfaces~\cite{fwm+11,cfgw11,lbs+17}. Moreover, these techniques have often been deployed as modifications to Android's permission enforcement~\cite{eom09b,egc+10,nkz10,fbj+12,dsp+11,fwm+11,bdd+12,omem09,cnc10,zzjf11,sc13,bdd+11b,hnes14,pfnw12,sdw12}, SDK tools~\cite{fch+11,azhl12,vcc11}, or inline reference monitors~\cite{xsa12,jmv+12,Davis2012IARMDroidAR,bgh+13,bbh+15}.   While this paper demonstrates the evaluation of only a small subset of these tools with \tool, our experiments demonstrate that \tool has the potential to impact nearly all of them.
For instance, we can apply \tool to
vet SSL analysis tools by purposely introducing complex SSL errors in
applications, or privilege or permission misuse analysis tools, by developing security operators that attempt to misuse permissions.

\section{Limitations}
\label{sec:limitations}

\myparagraph{1) Soundness of \tool} As acknowledged in  Section~\ref{sec:relwork}, \tool{} does not aim to supplant formal verification (which would be sound) and does not claim soundness guarantees. Rather, \tool{} provides a systematic approach to semi-automatically uncover flaws in existing security tools, which is a significant advancement over manually-curated tests.

\myparagraph{2) Manual Effort} Presently, the workflow of \tool requires an  analyst to manually analyze the result of \tool (\ie, uncaught mutants).
However, as described in Section~\ref{sec:executing_tools},~\tool possesses enhancements that mitigate the manual effort by dynamically eliminating non-executable mutants\remove{,} that would otherwise impose a burden on the analyst examining undetected mutants. In our experience, this analysis was completed in a reasonable time using the methodology outlined in Section~\ref{sec:feasibility}.

\myparagraph{3) Limitations of Execution Engine} Like any dynamic analysis tool,
the EE will not explore all possible program states, thus, there may be a set of mutants marked as non-executable by the EE, that may actually be executable under certain scenarios.  However, the {\sc \small CrashScope} tool, which \tools EE is based upon, has been shown to perform comparably to other tools in terms of coverage~\cite{Moran:ICST16}.
Future versions of \tool's EE could rely on emerging input generation tools for Android apps~\cite{Mao:ISSTA16}.

\myparagraph{4) Dependency on Android Framework APIs} We designed \tool{} to be as generic as possible, as discussed in Section~\ref{sec:scheme-generality} and Section~\ref{sec:implementation}.
For example, the mutation seeding methodology relies on the AST of the target source code that \textit{selects} the target location for mutation.
As a result, as long as the Android framework changes through extension and target code are parse-able as AST, the seeding methodology will not have to be changed.
Indeed, the base apps we used for \tool{} (Table~\ref{tbl:app_list} in the Appendix) rely on different versions of Android SDK with both Gradle~\cite{gradle} and pre-Gradle build system, which demonstrates the \textit{versatility} of \tool{}.
On the other hand, calls to functions/methods from the Android framework APIs are introduced through mutation.
Therefore, these \textit{will} have to be changed as Android Framework changes over time, as it is \textit{not possible} to generalize such calls.
% }

% {\color{blue}
\myparagraph{5) Adaptation to Different Goals} \tool requires defining security operator through manual examination of the claims made by existing tools \ie, the security goal of the concerned tool
(Section~\ref{sec:operators}).
Further changes might be necessary for satisfying syntactical requirements related to the defined, new, security operator.
For example,
for \texttt{Cipher.getInstance}, the security operator will have to be enclosed within a try-catch scope because of its \texttt{throws-exception}
% method
signature.
Thus, the implementation of security operators to mutate cryptographic APIs may require manual intervention by domain experts initially, but we expect it to be a one-time effort,
similar to how we defined security operators for data leak APIs once for detecting flaws in this paper.
Finally, we note that \tools modular implementation separates the operators from its core components, \ie the Mutation Engine (ME) and the Execution Engine (EE) simply seed and execute security operators as per the operator specification, and hence, are decoupled from the security goal per se.
Hence, the
implementation of
\tool framework would not have to change for different security goals, limiting the amount of code change to only the addition of
goal-specific security operators.
% }

\section{Conclusion}
\label{sec:conc}
We proposed the \tool framework for performing systematic security evaluation of Android static analysis tools to discover (undocumented) unsound assumptions, adopting the practice of mutation testing from SE to security. \tool not only detected major flaws in popular, open-source Android security tools, but also demonstrated how these flaws propagated to other tools that inherited the security tool
or followed similar principles. With \tool, we demonstrated how mutation analysis can be feasibly used for gleaning unsound assumptions in existing tools, benefiting developers, researchers, and end users, by making such tools more secure and transparent.

\vspace{-0.1em}
\begin{acks}
%\section{Acknowledgements}
We thank the developers of the evaluated tools for making their tools available
to the community, and for being open to suggestions. We thank Richie Bonnett for his contributions to the conference paper version of this work.
This work is supported
in part by the NSF-1815336 and NSF-1815186.
\end{acks}

\bibliographystyle{ACM-Reference-Format}
\bibliography{bib/os,bib/taint,bib/phone,bib/misc,bib/semeru,bib/mutation-references,bib/pl,bib/mutation_security,bib/evaluation}

\appendix

\section*{Appendix}
\begin{lstlisting}[basicstyle=\ttfamily\scriptsize,caption={\small A
dynamically-created BroadcastReceiver, created inside another, with data leak. Whenever the {\sf onReceive()} callback of the {\sf receiver} object is invoked, it will create another {\sf receiver} object of similar type, with a leak inside its own {\sf onReceive()} callback. This can be further evolved to use anonymous object declaration that can leak information in a similar nature.
    },label=lst:inception,emph={},emphstyle=\bfseries]
    BroadcastReceiver receiver = new BroadcastReceiver() {
        @Override
        public void onReceive(Context context, Intent intent) {
        BroadcastReceiver receiver = new BroadcastReceiver(){
            @Override
                public void onReceive(Context context, Intent intent) {
                    String dataLeak = Calendar.getInstance().getTimeZone().getDisplayName();
                    Log.d("leak-1", dataLeak);}};
        registerReceiver(receiver, new IntentFilter().addAction("android.intent.action.SEND"));
    }};
    registerReceiver(receiver, new IntentFilter().addAction("android.intent.action.SEND"));
    \end{lstlisting}

\myparagraph{CrashScope (Execution Engine)} The EE functions builds upon CrashScope~\cite{Moran:ICST16, Moran:ICSE17}, which statically analyzes the code of a target app to identify  activities implementing potential contextual features (\eg{} rotation, sensor usage) via API call-chain propagation.
This execution is guided by one of several exploration strategies, organized along three dimensions: (i) GUI-exploration, (ii) text-entry, and (iii) contextual features.

Note that because the goal of the EE is to explore as many screens of a target app as possible, the EE forgoes certain combinations of exploration strategies from {\sc \small CrashScope}~\cite{Moran:ICST16, Moran:ICSE17} (\eg{} entering unexpected text or disabling contextual features) prone to eliciting crashes from apps.
The approach uses \texttt{adb} and Android's \texttt{uiautomator} framework to interact with and extract GUI-related information from a target device or emulator.
Further implementation details of exploration strategies can be found in \cite{Moran:ICST16, Moran:ICSE17}.

\begin{table*}[ht]
    \centering
    \scriptsize

    \caption{{\small List of App names, URLs and IDs assigned by us for the purpose of the \tool study}}
        \begin{tabular*}{\textwidth}{l @{\extracolsep{\fill}} l l}
        \toprule
        \textbf{App ID} & \textbf{Andorid App name} & \textbf{URL}                                                            \\
        \midrule
        app 01 & 2048                      & \url{https://f-droid.org/en/packages/com.uberspot.a2048/}                     \\
        app 02 & Protect Baby Monitor      & \url{https://f-droid.org/en/packages/protect.babymonitor/}                    \\
        app 03 & QR Scanner                & \url{https://f-droid.org/en/packages/com.secuso.privacyFriendlyCodeScanner/}  \\
        app 04 & Location Share            & \url{https://f-droid.org/en/packages/ca.cmetcalfe.locationshare/ }            \\
        app 05 & Camera Roll               & \url{https://f-droid.org/en/packages/us.koller.cameraroll/ }                  \\
        app 06 & AndroidPN Client          & \url{https://f-droid.org/en/packages/org.androidpn.client/  }                 \\
        app 07 & Activity Launcher         & \url{https://f-droid.org/en/packages/de.szalkowski.activitylauncher/}         \\
        app 08 & Man Man                   & \url{https://f-droid.org/en/packages/com.adonai.manman/}                      \\
        app 09 & BMI Calculator            & \url{https://f-droid.org/en/packages/com.zola.bmi/}                           \\
        app 10 & A Time Tracker            & \url{https://f-droid.org/en/packages/com.markuspage.android.atimetracker/}    \\
        app 11 & AFH Downloader            & \url{https://f-droid.org/en/packages/org.afhdownloader/}                     \\
        app 12 & Android Explorer          & \url{https://f-droid.org/en/packages/com.iamtrk.androidexplorer/}             \\
        app 13 & Kaltura Device Info       & \url{https://f-droid.org/en/packages/com.oF2pks.kalturadeviceinfos/}          \\
        app 14 & Apod Classic              & \url{https://f-droid.org/en/packages/com.jvillalba.apod.classic/}             \\
        app 15 & Calendar Trigger          & \url{https://f-droid.org/en/packages/uk.co.yahoo.p1rpp.calendartrigger/}      \\
        \bottomrule
        \end{tabular*}\label{tbl:app_list}
\end{table*}

\myparagraph{Apps used in the study} For our study, we collected a set of 15 open-source apps from F-Droid~\cite{fdroid}, as shown in \autoref{tbl:app_list}.
The apps come from different heterogeneous build configuration settings, with compile SDK API level 23 (Marshmallow) to 27 (Oreo), minimum SDK API level 9 (Gingerbread) to 16 (Jelly Bean), and target SDK API level from 17 (Jelly Bean) to 26 (Oreo), with sizes ranging from several hundred KB to a maximum of 3.5 MB.

\end{document}